\documentclass[superscriptaddress,amsmath,amssymb, aps, prx,longbibliography,twocolumn, nofootinbib]{revtex4-2} 
\usepackage{upgreek}
\usepackage{float}
\usepackage{ulem}

\usepackage{xfrac}
\usepackage{color}
\usepackage{graphicx}
\usepackage{nccmath}
\usepackage{bm}
\usepackage{hyperref}
\hypersetup{colorlinks,breaklinks,
            urlcolor=[rgb]{0,0,0.36},
            linkcolor=[rgb]{0,0,0.36},
            citecolor=[rgb]{0,0,0.36}}
\usepackage[mathlines]{lineno}
\usepackage{dcolumn}
\usepackage{mathtools}  
\usepackage{siunitx}    
\usepackage{dsfont}
\usepackage{CJKutf8}
\usepackage[toc,page]{appendix}

\usepackage{epsfig,tabularx,multirow,booktabs}

\usepackage{braket}

\usepackage[dvipsnames]{xcolor}
\definecolor{cadmiumgreen}{rgb}{0.0, 0.42, 0.24} 

\usepackage{pifont} 

\renewcommand{\v}[1]{\boldsymbol{#1}}

\usepackage{adjustbox} 
\usepackage{tabularx} 
\newcolumntype{Y}{>{\centering\arraybackslash}X} 

\makeatletter
\newcommand{\suppressTOC}{%
  \let\orig@addcontentsline\addcontentsline
  \renewcommand{\addcontentsline}[3]{}%
}
\newcommand{\restoreTOC}{%
  \let\addcontentsline\orig@addcontentsline
}
\makeatother
\suppressTOC

\makeatletter
\renewcommand\footnotesize{%
   \@setfontsize\footnotesize\@ixpt{11}%
   \abovedisplayskip 1\p@ \@plus2\p@ \@minus4\p@
   \abovedisplayshortskip \z@ \@plus\p@
   \belowdisplayshortskip 2\p@ \@plus2\p@ \@minus2\p@
   \def\@listi{\leftmargin\leftmargini
               \topsep 2\p@ \@plus2\p@ \@minus2\p@
               \parsep 1\p@ \@plus\p@ \@minus\p@
               \itemsep \parsep}%
   \belowdisplayskip \abovedisplayskip
}
\makeatother

\newcommand{\Harvard}{Department of Physics, Harvard University, Cambridge, Massachusetts 02138, USA}

\newcommand{\Berkeley}{Department of Physics, University of California, Berkeley, CA 94720, USA}

\newcommand{\UPenn}{Department of Physics and Astronomy, University of Pennsylvania, PA 19104, USA}

\newcommand{\LBNL}{Material Science Division, Lawrence Berkeley National Laboratory, Berkeley, CA 94720, USA}

\begin{document}

\title{Robust superconductivity upon doping chiral spin liquid and Chern insulators in a Hubbard-Hofstadter model}

\author{Clemens Kuhlenkamp}
\thanks{These authors contributed equally to this work.\\}
\affiliation{\Harvard}

\author{Stefan Divic}
\thanks{These authors contributed equally to this work.\\}
\affiliation{\UPenn}
\affiliation{\Berkeley}

\author{Michael P. Zaletel}
\affiliation{\Berkeley}
\affiliation{\LBNL}

\author{Tomohiro Soejima (\begin{CJK*}{UTF8}{bsmi}副島智大\end{CJK*})}
\affiliation{\Harvard}

\author{Ashvin Vishwanath}
\affiliation{\Harvard}

\date{\today}

\begin{abstract}
Demonstrating superconductivity in purely repulsive Hubbard models is a compelling goal which underscores the counter-intuitive ability of Coulomb interactions to mediate superconductivity.
Here, we present numerical evidence for robust superconductivity in a triangular Hubbard-Hofstadter model at $\pi/2$ flux per plaquette. Employing infinite density matrix renormalization group calculations on infinite cylinders of finite circumference, we observe superconducting ground states for a wide range of dopings, whose pair-correlations strengthen as the 2D limit is approached. 
At a density of one electron per site, Hubbard interactions have been reported to drive the insulating parent state of the superconductor from an integer quantum Hall (IQH) state to a chiral spin liquid (CSL). Our findings give credence to a recent proposal that proximity to the IQH-CSL transition serves to make chiral superconductivity energetically favorable on doping, and also correctly predicts the nature of the edge modes in the superconductor. On the CSL side, this suggests the superconductor can be thought of as arising from Laughlin's `anyon superconductivity' mechanism.
Thus the Hubbard-Hofstadter model studied here offers a clean and experimentally accessible setup --- potentially realizable in moir\'e heterostructures --- for exploring the properties of anyonic matter at finite density and the interplay of topological order, quantum criticality and superconductivity.
\end{abstract}

\maketitle

\section{Introduction}

Doped charges enter gapped topologically-ordered phases such as fractional Chern insulators (FCIs) and spin liquids by fractionalizing into anyons. The fate of such phases upon doping a finite density of charge then relies on the properties of mobile anyon gases. The resulting collective behaviors are typically hard to ascertain, as the parent states themselves are already strongly correlated phases of matter.
The experimental observation of FCIs in two-dimensional materials~\cite{Cai_SignaturesFractional_2023,Zeng_ThermodynamicEvidence_2023,Park_ObservationFractionally_2023,Xu_ObservationInteger_2023,LongJu2024}, often in proximity to superconductivity~\cite{LongJuSc2025,xu2025_sc_mote2,Nguyen2025}, has sparked renewed interest in this broad question.

Remarkably, early theoretical work in the context of high-$T_c$ superconductors proposed that certain anyons, such as charged semions in chiral spin liquids (CSLs)~\cite{Kalmeyer87}, can naturally give rise to unconventional superconductivity at low temperatures~\cite{LaughlinRelationshipHighTemperatureSuperconductivity1988, LaughlinSuperconductingGroundState1988, FetterRandomphaseApproximationFractionalstatistics1989, LeeAnyonSuperconductivityFractional1989, WenChiralSpinStates1989, HosotaniSuperconductivityAnyonModel1990, WenCompressibilitySuperfluidityFractionalstatistics1990, LeeAnyonSuperconductivity1991,ChenAnyonSuperconductivity1989,Song2021}.
While CSL ground states have been identified in repulsive systems~\cite{he2014chiral, gong2014emergent,gong2015global, bauer2014chiral, szasz2020chiral, chen2022quantum, wietek2021mott}, superconductivity upon doping is often suppressed in favor of competing phases such as chiral metals~\cite{jiang2017holon, peng2021doping, zhu2022doped}.

Additional energetic input is clearly required in order to justify the robust formation of `anyon superconductivity'. Recent theoretical work has shown that the hierarchy of anyon excitation energies is a key input for determining the doped phase~\cite{Shi2024,DivicPNAS,Pichler2025}. In a CSL, for instance, a charge-$e$ semion must be cheaper than adding a bare electron. One broadly applicable strategy identified in Ref.~\cite{DivicPNAS} is to tune the system to an insulator-insulator quantum critical point where the desired anyon energy gap softens. Doping in the vicinity of this topological critical point should then yield a dilute gas of the relevant anyons, setting the stage for anyon superconductivity. While a combination of analytical and numerical studies have begun to explore this and other mechanisms~\cite{jiang2020topologicalPRL,Jiang2021, huang2022topological, Jiang2023, huang2023quantum,DivicPNAS,Zhang2025,Shi2025, Nosov2025,Pichler2025,Guerci2025,Gattu2025,Xu2025clusters,Han2025}, direct numerical evidence of superconductivity in purely electronic models at finite doping remains elusive. 

Lattice models exhibiting such critical behavior have been constructed for both the CSL and a $\nu= 2/3$ fractional Chern insulator \cite{Kuhlenkamp24,Divic24,Pichler2025}. Nevertheless, these earlier numerical studies were confined to zero doping, so the possibility of competing non-superconducting phases at finite carrier density could not be excluded. Here we remedy this limitation: we report on large-scale simulations performed directly at finite doping that unambiguously show the resulting ground state is superconducting.

Concretely, we study a Hubbard-Hofstadter model on the triangular lattice, differing from the triangular lattice Hubbard model only by the insertion of $\pi/2$ flux through each triangular plaquette. This modification allows for a non-interacting band insulator, specifically a Chern insulator, at a filling of one electron per site, which transitions into a CSL on increasing on-site repulsion~\cite{Kuhlenkamp24,Divic24}. Close to the insulator-insulator transition, where the Hall conductance changes, all spin-carrying excitations remain gapped while spinless charge excitations, including charge-$e$ semions, are pushed down in energy~\cite{DivicPNAS}. We investigate this model in this regime, at finite doping, using infinite matrix renormalization group (iDMRG) simulations on cylinders of finite width and infinite length. In this setting, we numerically demonstrate that chiral superconductivity out-competes other states, such as metals or charge density waves.

The robust superconductivity and absence of competing phases in our simulations stand in stark contrast to the square and triangular-lattice Hubbard models at zero flux, where the relation between doped and parent states is often difficult to establish~\cite{ArovasAnnRev2022,zhu2022doped}. 
Our results demonstrate that the Hubbard-Hofstadter model provides a minimal, clean platform for superconductivity emerging from both a spin liquid and quantum Hall insulator, with the prospect of being realized in experiments~\cite{Kuhlenkamp24}.

\begin{figure}
\centering
\includegraphics[width=\linewidth]{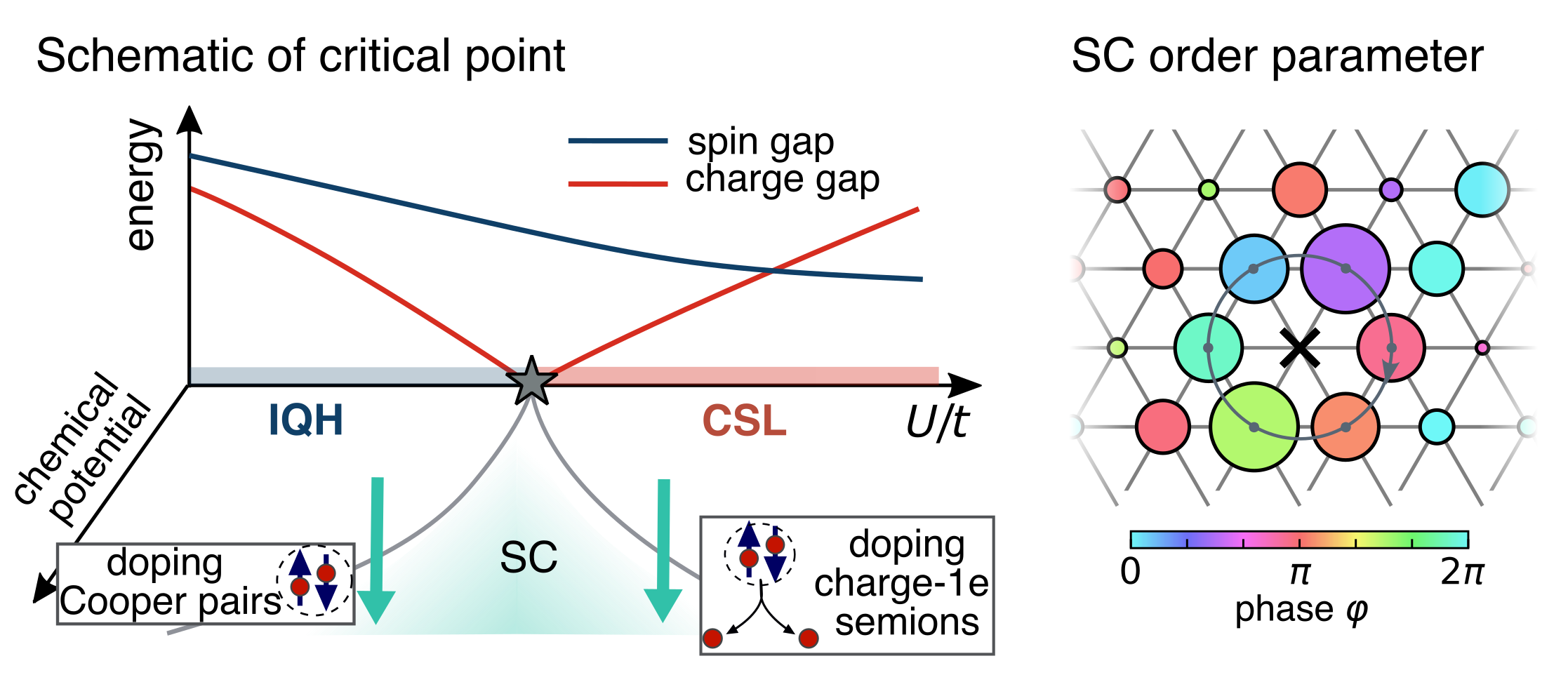}
\caption{\textbf{Superconductivity emerging near topological criticality.} Left panel: 
Proposed schematic phase diagram of the Hubbard-Hofstadter model at $\phi_\triangle =\pi/2$ flux per triangle. On-site repulsion $U$ drives a Mott transition between integer quantum Hall (IQH) and chiral spin liquid (CSL) phases, indicated by the gray star. The change in Hall conductivity from $\sigma_{xy}=2e^2/h$ to $0$ requires the charge gap (red curve) to vanish at a continuous transition, while the spin gap remains open (blue curve); charge-$e$ semions and Cooper pairs are then the fundamental low-lying excitations of the CSL and IQH phases, respectively, near the transition. Doping each phase leads to the same chiral superconductor (SC). Right panel: Numerical proxy for superconducting order parameter in the doped CSL, demonstrating site-centered $p$-wave pairing symmetry. Computed on the YC4 cylinder at Hubbard strength $U/t=14$, hole doping $\delta = 1/8$, and MPS bond dimension $\chi=6144$.}
\label{fig:mainmessage}
\end{figure}

\section{Model and phase transition}

We study the triangular lattice Hubbard model in an orbital magnetic field $B_z$, without Zeeman coupling:
\begin{align}
    \label{eq:microscopic-model}
    \hat{H} &= - t \sum_{\langle ij\rangle,\sigma} e^{i \phi_{ij}} \, c^\dagger_{i,\sigma} c_{j,\sigma} + \mathrm{h.c.}
    + U \sum_i n_{i,\uparrow} \, n_{i,\downarrow},
\end{align}
where $\sigma =\, \uparrow,\downarrow$ is a (pseudo-)spin degree of freedom. The phases $\phi_{ij}$ are chosen such that flux $\Phi_{\triangle} = B_z a^2 \sqrt{3}/4$ permeates each plaquette, where $a$ is the lattice constant.

We choose $\phi_{\triangle}  = 2\pi\Phi_{\triangle}/\Phi_0 = {\pi}/{2}$ flux through every triangle, with $\Phi_0 = hc/e$. Two features of this $\pi/2$ flux model must be emphasized. First, the magnetic unit cell consists of two sites, allowing for a band insulator even at `half filling', \textit{i.e.}, density of one electron per site. Second, this model enjoys particle-hole symmetry \cite{Yang1989,1990YangZhang,Yang1991, Affleck1990, DivicPNAS}, so that hole doping and electron doping away from half filling are equivalent.
At half filling and small $U/t$, the ground state is a spin-singlet integer quantum Hall (IQH) state with Hall conductance $\sigma_{xy}=2{e^2}/{h}$. Increasing $U$, the system enters a Mott insulating phase with robust CSL topological order~\cite{Kuhlenkamp24,Divic24}.
This CSL persists up to $U/t \sim 20$, before the system eventually orders magnetically~\cite{Gong2017,Wietek2017,Kuhlenkamp24}, most likely into the \SI{120}{\degree} coplanar antiferromagnet~\cite{Miyashita1984, huse1988simple}. Consistent with this observation, the effective three-spin chiral coupling~\cite{sen1995chiral} at this value of $t/U$ falls within the range where its competition with two-spin super-exchange leads to magnetic order in the effective Heisenberg model~\cite{Kuhlenkamp24, Hu2016, motrunich2006spinliquid, Wietek2017, Gong2017, Saadatmand2017, Huang2024}.

We briefly review the proposed theory of the IQH-CSL transition~\cite{DivicPNAS,KuhlenkampThesis,Zhou_deconf_2025} and the route to superconductivity~\cite{DivicPNAS}. Both the IQH phase at weak coupling and the CSL phase at intermediate-to-strong coupling are spin $\mathrm{SU(2)}$ symmetric and share the same charge-neutral edge modes. Previous studies have proposed that there can be a continuous phase transition between them, which is consistent with numerical studies of Eq.~\eqref{eq:microscopic-model} at $\phi_\triangle=\pi/2$~\cite{Kuhlenkamp24, Divic24}, and in related continuum models~\cite{Zhou_deconf_2025}.
Due to the change in Hall conductance between these phases, the charge gap must close if the transition is continuous,
while the spin gap can remain open; see Fig.~\ref{fig:mainmessage}(a) for a schematic depiction of doping the system close to the IQH to CSL phase transition~\cite{DivicPNAS,KuhlenkampThesis}.

On the IQH side, the spin-singlet Cooper pair is the cheapest charge excitation close to the critical point, as it avoids the spin gap. Thus superconductivity can naturally arise by condensing these pairs. Further, one can show that the topology of the IQH phase, namely its gapless spin edge mode and associated spin quantum Hall response, is inherited by the superconductor (see App.~\ref{app:field_theory}).

Doping the CSL is more subtle due to its fractionalized bulk excitations. An electronically-realized CSL harbors two anyons distinguished by their quantum numbers: the usual `spinful semion' which carries $(Q,s)=(0e,\hbar/2)$ relative to the ground state, and a `charged semion' with $(Q,s)=(1e,0\hbar)$, their composite being the electron.
Close to the IQH-CSL transition, the charged semion is the lowest-energy excitation of the CSL and may be interpreted as a fractionalized spin-singlet Cooper pair. Therefore, Ref.~\cite{DivicPNAS} proposed that the $\phi_\triangle = \pi/2$ Hubbard-Hofstadter model provides at finite doping the ideal energetic conditions for the microscopic realization of Laughlin's anyon superconductivity proposal~\cite{LaughlinRelationshipHighTemperatureSuperconductivity1988,LaughlinSuperconductingGroundState1988}.

\section{Numerical methods}

\begin{figure}
\centering
\includegraphics[width=\linewidth]{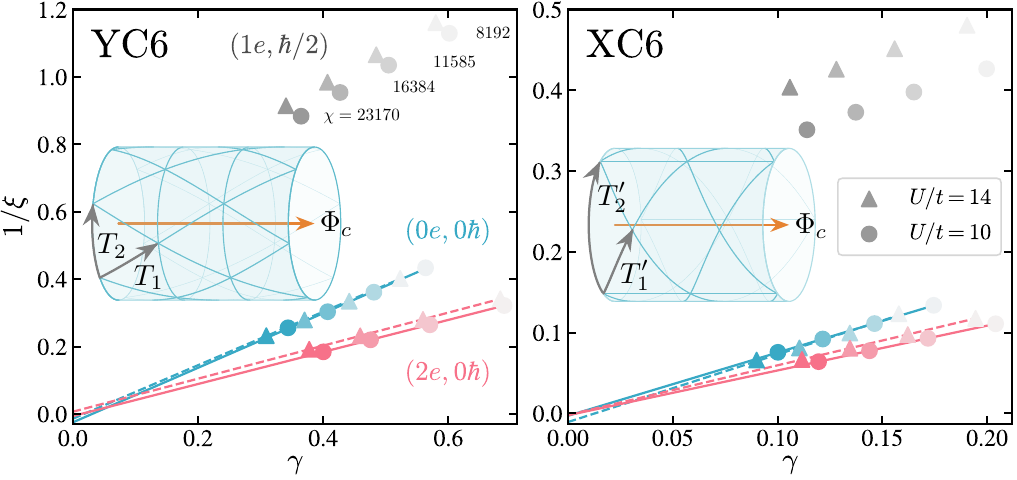}
\caption{\textbf{Scaling of correlation lengths} with a parameterization $\gamma$ of the finite MPS bond dimension in the $0e,1e,2e$ charge sectors for cylinder geometries YC6 at doping $\delta=1/12$ (left panel) and XC6 at $\delta=1/9$ (right panel), at $U/t=10$ (circles) and $14$ (triangles). In each charge sector, $\gamma$ is defined in terms of the difference between the largest transfer matrix eigenvalues as described in the main text.
Linear fits in the $0e$ and $2e$ sectors at both $U/t=10$ (solid lines) and $14$ (dotted lines).
Insets: Illustrated segments of the YC6 and XC6 cylinders threaded by $\Phi_c$ charge flux (represented by orange arrows), with translation operators shown.
}
\label{fig:scaling_xis}
\end{figure}

We study the Hubbard-Hofstadter model at $\phi_\triangle = \pi/2$ compactified to infinite-length cylinders of finite circumference. We focus on two families of geometries, namely the YC-$L_y$ and XC-$2L_y$ cylinder geometries~\cite{szasz2020chiral}; see insets in Fig.~\ref{fig:scaling_xis}.
We use $T_x$ to denote $T_1$ ($T_1'$), and likewise $T_y$ to denote $T_2$ ($T_2'$), on the YC (XC) cylinders. We label the lattice sites by coordinates $(x,y)\in \mathbb{Z}\times \mathbb{Z}_{L_y}$ such that acting with $T_{x/y}$ increments $x/y$ by unity. We work in a hybrid position-momentum basis defined by $
c^\dag(x,k_y) =  L_y^{-1/2} \sum_{y=0}^{L_y-1} e^{-ik_y y} c^\dag({x,y}) $,
which carries eigenvalue $e^{ik_y}$ under $T_y$.

Threading flux $\Phi_c$ through a given cylinder twists the boundary condition according to $T_y^{L_y} = e^{i\phi_c N_F}$, where $N_F$ is the fermion number operator and $\phi_c = 2\pi \Phi_c / \Phi_0$. The eigenvalues $e^{ik_y}$ of $T_y$ in the charge-$N_F$ sector correspond to circumferential momenta $k_y = ({2\pi m + N_F\phi_c})/L_y$, where $m \in \{0, 1, \dots, L_y-1\}$.

Choosing $\phi_c$ dictates the symmetries of the cylinder system~\cite{Divic24}.
The IQH-CSL transition was identified as having a particle-hole (PH) symmetry~\cite{DivicPNAS,KuhlenkampThesis}. Hence, we choose $\phi_c$ on each cylinder
so that the system is PH-symmetric before doping (see App.~\ref{app:lattice_conventions}). We focus on small hole doping, $\delta \in [1/15,\, 1/8]$, corresponding to an average density of $n = 1-\delta$ electrons per site; the scenario of electron doping is related by the PH transformation. All simulations were performed using the TeNPy library~\cite{tenpy, tenpy_v1}. Details of our iDMRG numerics are provided in App.~\ref{app:comp_details}.

\section{Charge doping close to criticality}
\label{sec:evidence_for_SC}

Here we detail evidence for superconductivity upon doping the IQH and CSL states near their phase boundary at approximately $U/t \approx 11 \sim 12$~\cite{Kuhlenkamp24,Divic24}. We focus on the interaction strengths $U/t=10$ (IQH) and $14$ (CSL). To capture the nature of the ground states, we analyze correlation functions of the ring-averaged density $n(x)$, spin density $S_z(x)$, momentum-resolved Cooper pair operator $\Delta_d(x, P_y)$, and the electron creation operator $c^\dagger(x, k_y)$. We define\footnote{For the electron creation operator, the correlation function $C_{c^\dagger}$ further sums over momentum, as described in App.~\ref{app:def_corr_functions}.} $C_{\mathcal{O}}(x) = \langle \mathcal{O}^\dagger(x) \mathcal{O}(0) \rangle - \langle \mathcal{O}^\dagger(x) \rangle \langle \mathcal{O}(0) \rangle$. Here $d$ is the orientation of the Cooper pair (see App.~\ref{app:def_corr_functions} for the definitions of these operators).

Due to the quasi-1D nature of the studied cylindrical geometries, the $\mathrm{U}_c(1)$ symmetry is unbroken in the ground state~\cite{hohenberg1967existence, mermin1966absence, coleman1973there,takada1975long}, with superconductivity manifesting as quasi long-range order (QLRO) of pair correlations:
\begin{equation}
  C_{\Delta}(x) \sim {x^{-K_\text{SC}}} + \dots
\end{equation}
This defines the superconducting exponent $K_\mathrm{SC}$. In stark contrast to metals, 
quasi-1D superconductors are gapped to single-electron and spin excitations~\cite{LutherEmery_PRL1974}, which are associated with correlations that decay exponentially. 

From bosonization considerations, these long-range superconducting correlations are accompanied by power-law decaying correlations for the density operator $n$:
\begin{equation}
  C_{n}(x) \sim  x^{-2} + A \,{x^{-K_\mathrm{CDW}}} \cos(2\pi x\,\rho_{\text{pair}}^\mathrm{1D}) +\dots, 
  \label{eq:K_CDW}
\end{equation} 
where the first term is a universal contribution due to $\mathrm{U}_c(1)$ charge conservation, and the second term oscillates with a wavelength governed by the inverse density of Cooper pairs $1/\rho_{\text{pair}}^{\text{1D}} = 2/(L_y \delta)$~\cite{Haldane1981}.
The two exponents are related by the Luther-Emery relation $K_\mathrm{SC} K_\mathrm{CDW} = 1$, which highlights that enhancing superconducting correlations suppresses charge density wave fluctuations.

For states represented as an iMPS, the correlation function of any operator $\mathcal{O}$ of charge $c$ decomposes as
\begin{align} \label{eq:corr_decomposition}
    C_{\mathcal{O}}(x) = \sum_i o_i \lambda_{c, i}^x,
\end{align}
where $\lambda_{c,i}$ are the transfer matrix eigenvalues in charge sector $c$, and $o_i$ are form factors~\cite{schollwock2011density}. For large $x$, Eq.~\eqref{eq:corr_decomposition} is dominated by the eigenvalue of largest magnitude:
\begin{align} \label{eq:def_xi_charge}
    \sum_i o_i \lambda_{c, i}^x \sim o_1 \lambda_{c, 1}^x \equiv o_1 e^{i\alpha x} e^{-{x}/{\xi_c}},
\end{align}
We call $\xi_c$ the `correlation length' in the charge sector $c$, which we report in units of ring distance $x$.

When the correlation function decays as a power-law at long distances, the correlation length $\xi_c$ must diverge as $\chi \to \infty$. To analyze this divergence, we define a scaling parameter $\gamma$ in each sector, which parametrizes the distance between the two dominant transfer matrix eigenvalues, following Refs.~\cite{rams2018precise,Vanhecke2019scaling} (where the scaling parameter is denoted by $\delta$).
Fig.~\ref{fig:scaling_xis} shows $1/\xi$ versus $\gamma$ in the $0e$, $1e$, and $2e$ charge sectors on the YC6 and XC6 cylinders. As $\gamma \to 0$ with increasing MPS bond dimension, the $0e$ and $2e$ correlation lengths diverge in proportion to $1/\gamma$ (established by linear fits), whereas the $1e$ correlation length remains finite, consistent with a Luther-Emery liquid but not a spinful Luttinger liquid~\cite{LutherEmery_PRL1974}.

The momentum sector labeling each $\xi$ further characterizes the low-lying excitations. In the charge-$0e$ sector, the diverging correlation length corresponds to $k_y = 0$, consistent with the momentum of the ring-averaged density operator $n(x)$. In contrast, in the charge-$2e$ sector the dominant momentum is $k_y = \pi$ on all studied cylinder geometries, so that
the fluctuating Cooper pairs carry eigenvalue $T_y = -1$, as predicted in Ref.~\cite{DivicPNAS}.

\begin{figure}
\centering
\includegraphics[width=\linewidth]{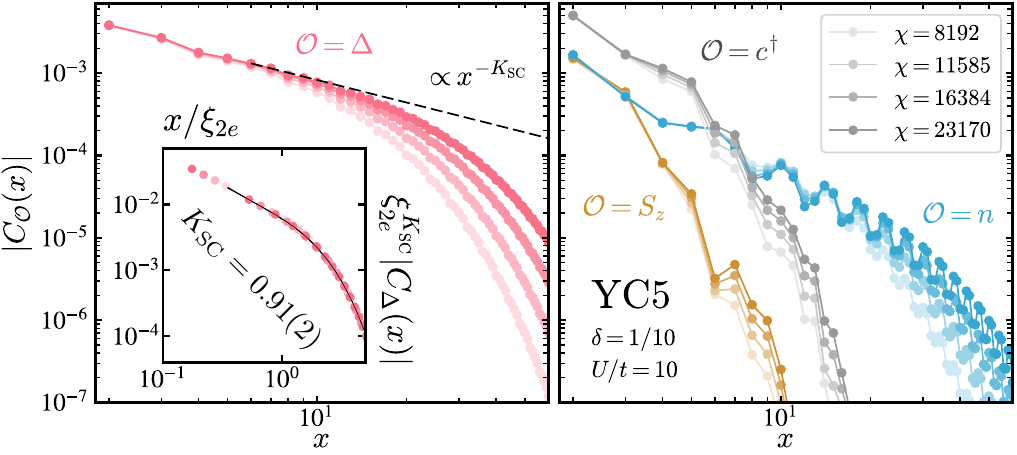}
\caption{\textbf{Dominant fluctuations in the doped state.} Left panel: Magnitude of pair correlation function vs. distance (in units of rings) along the infinite YC5 cylinder, at $\delta=1/10$ and $U/t=10$.
Shade darkens with increasing MPS bond dimension indicated in legend.
Inset: Optimal scaling collapse of pair correlations, achieved with an exponent $K_\text{SC}=0.91\pm0.02$. The determined power law is shown in the main figure with a dashed line. Here we took  $d=1, k_y = \pi$ for $\Delta$. Right panel: Magnitude of connected correlations of the ring-averaged charge density (blue), spin density (orange), and electron operator (gray) for the same parameters.
}
\label{fig:four_corrs}
\end{figure}

We further analyze the correlations $C_{\mathcal{O}}$. Across system sizes and hole dopings, we find that superconducting correlation $C_\Delta$ with $k_y = \pi$ approach a power-law with increasing $\chi$, as shown in Fig.~\ref{fig:four_corrs}(a) for the YC5 geometry at $\delta=1/10$ (see App.~\ref{eq:app_correlation_data} for correlations in all systems). Density correlations tend towards an oscillatory power law (Fig.~\ref{fig:four_corrs}b), with a wavelength indeed equal to $1/\rho^\mathrm{1D}_\mathrm{pair}$. In contrast, spin and electron correlations show rapid exponential decay, consistent with a Luther-Emery liquid.

We remark that we found no evidence of discrete symmetry breaking on the studied geometries. Namely, while finite $\chi$ induces a small spatial variation in $\langle n(x) \rangle$, the density becomes uniform as $\chi$ increases; see App.~\ref{app:comp_details}.

Superconductivity prevails in the 2D limit if pair correlations become long-ranged, $K_\mathrm{SC}\rightarrow 0$ as  $L_y\rightarrow \infty$~\cite{Gannot2023}. In our simulations, QLRO at finite circumference is cut off by the finite MPS bond dimension $\chi$, and is recovered as $\chi\rightarrow \infty$. We extract the power-law exponent $K_\mathrm{SC}$ by a scaling analysis~\cite{Sahay23}, where we postulate a scaling form
\begin{equation}
    C_{\Delta}(x) = \xi_{2e}^{-K_\mathrm{SC}} F(x/\xi_{2e}).
\end{equation}
Here, $\xi_{2e}$ is the charge-$2e$ correlation length, rendered finite by finite $\chi$. Optimizing the scaling collapse of $C_\Delta(x)$ at several $\chi$ allows us to fit $K_\mathrm{SC}$, as exhibited in the inset of Fig.~\ref{fig:four_corrs} for the YC5 cylinder at $\delta=1/10$ and $U/t=10$. See App.~\ref{app:fitting_procedure} for details of the fitting procedure.

\begin{figure}
\centering
\includegraphics[width=\linewidth]{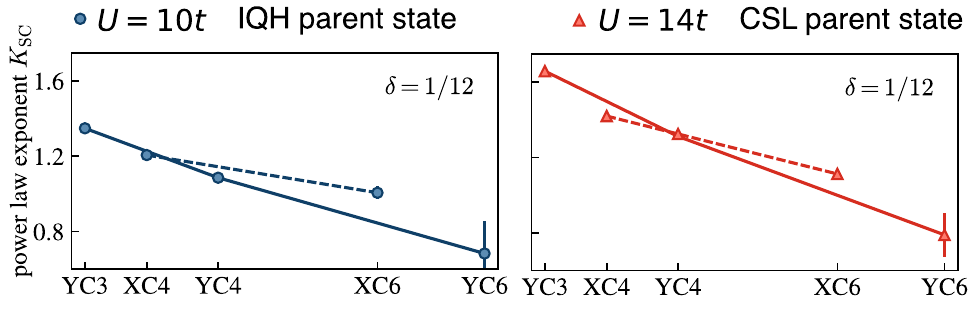}
\caption{\textbf{Evolution of exponents with circumference.}
Superconducting power-law exponents $K_\mathrm{SC}$ (minimized over directions $d$ with fixed pair momentum $k_y = \pi$) vs. cylinder circumference at $\delta=1/12$ for $U/t=10$ (left panel) and $U/t=14$ (right). Exponents are estimated by performing a scaling collapse as a function of MPS bond dimension; see inset of Fig.~\ref{fig:four_corrs}.
Values on the YC (XC) cylinder families are connected by solid (dashed) lines.
Within each family the exponent decreases monotonically with circumference, indicating the formation of superconductivity in the 2D limit.
}
\label{fig:K_SC_Lydependence}
\end{figure}

We demonstrate a clear trend: At each value of the hole doping, $K_\mathrm{SC}$ decreases monotonically with increasing cylinder circumference within each cylinder family at both $U/t=10$ and $14$, as shown in Fig.~\ref{fig:K_SC_Lydependence} for $\delta=1/12$.
Altogether, on both sides of the IQH-CSL transition, our findings are fully consistent with true 2D superconducting order compactified to a sequence of cylinders.

\section{Topological response and pairing symmetry}

\begin{figure}
\centering
\includegraphics[width=\linewidth]{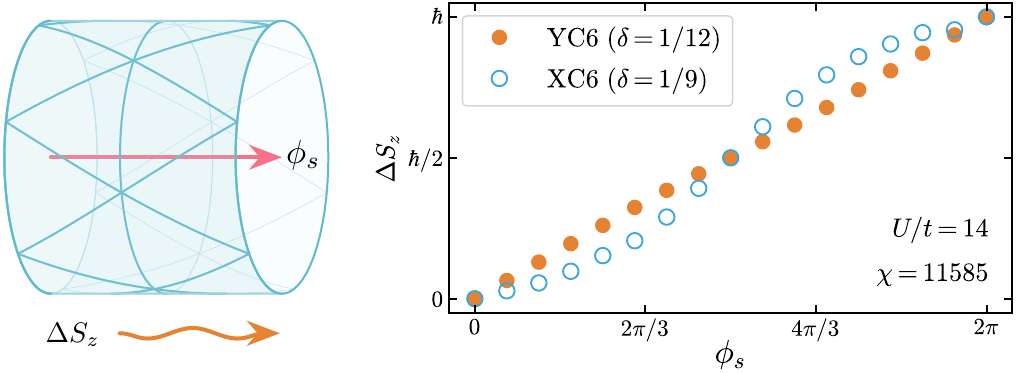}
\caption{\textbf{Topological response of superconductor.} Left panel: Illustration of cylinder segment where each ring is threaded by $\phi_s$ spin  flux, resulting in spin $\Delta S_z$ being pumped along the cylinder. Right panel: 
Pumped spin vs. threaded spin flux in the YC6 (XC6) cylinder geometry at $\delta = 1/12$ ($\delta = 1/9$) hole doping, plotted with orange disks (blue circles). Calculations performed at $U/t=14$ and $\chi=11585$.
}
\label{fig:spin_pump}
\end{figure}

Superconductors with a bulk spin gap exhibit a quantized spin quantum Hall conductance $\sigma_{xy}^s$, characterizing the linear response of spin current to a Zeeman field gradient~\cite{SenthilMarstonFisher}. We calculate this quantity by performing a Laughlin pumping simulation in the spin sector, namely by tracking the spin $\Delta S_z$ transported along the cylinder as a function of threaded spin flux $\phi_s$, as shown in Fig.~\ref{fig:spin_pump}, following a standard procedure (see App.~\ref{app:spin_response}). We obtain $\sigma^s_{xy} = 2(\hbar/2)^2 / h$, identical to that of the CSL and IQH.

This is predicted by the `anyon superconductivity' framework, wherein the superconductor inherits the quantized spin response of its parent state~\cite{Song2021,DivicPNAS}. A consistent but simpler picture emerges when doping Cooper pairs into the IQH phase; see Appendix~\ref{app:criticality_and_anyonSC}.

The quantized spin response obtained above is typically associated with the weak-pairing, topological $d+id$ superconductors~\cite{SenthilMarstonFisher}. However, the modification of the space group by the orbital magnetic field enforces a valley structure for electrons, which allows the superconductor to have site-centered $p$-wave pairing symmetry~\cite{DivicPNAS}.

To confirm this, we follow Ref.~\cite{Sahay23} and extract a proxy for $\Delta_{\v{0}}(\delta\mathbf{r}) =  \langle c(\delta\mathbf{r})\, is^y \, c(\v{0}) \rangle_\mathrm{2D}$, the 2D superconducting order parameter (see App.~\ref{app:pair_function}). We work in the $C_{6z}$ invariant gauge of Ref.~\cite{DivicPNAS}. The result is plotted in Fig.~\ref{fig:mainmessage}(b) for $\delta=1/8$ doping, which reveals that the superconducting pair wavefunction is odd under site-centered $C_{2z}$ rotations, \textit{i.e.}, $p$-wave.
Even though $C_{3z}$ symmetry is broken by the cylinder compactification, we clearly identify that the pair correlation function winds by $2\pi$ as $\delta\mathbf{r}$ traverses, clockwise, the six nearest neighbors of the origin.

\section{Discussion}

Having observed superconductivity upon doping consistent with the topological criticality mechanism~\cite{DivicPNAS}, we now examine how proximity to the putative quantum critical point dictates its properties. Starting from deep within the CSL or IQH parent phases at finite $U$, standard arguments predict a phase stiffness that tracks the doped charge density $\rho_s\sim\delta$~\cite{ChenAnyonSuperconductivity1989,DivicPNAS}. In contrast, near criticality the only length scale is the mean separation of doped holes $\ell \sim \delta^{-1/2}$; with dynamical critical exponent $z=1$ this gives the  `super-universal' relation $\rho_s\sim \ell^{-1} \sim \delta^{1/2}$, as recently emphasized for exciton anyon superfluids~\cite{Han2025}. The square-root scaling signals a much stiffer condensate, highlighting the practical advantage of doping near criticality and providing a clear prediction for experiments. We highlight the recent analysis in Ref.~\cite{Zhou_deconf_2025} of a related model, which provides strong evidence that the IQH-CSL transition can be continuous and described by a conformal field theory.

In the presence of long-range Coulomb repulsion or impurity pinning relevant to experimental realizations~\cite{Kuhlenkamp24}, the doped charges might crystallize rather than condense. On the CSL side, assuming the spin gap remains intact, the doped charge-$e$ semions can lock into a `semion crystal' (or CSL-CDW) with period determined by $1/\delta$, extending the CSL phase to finite density. On the adjacent IQH side the spin-gapped carriers are Cooper pairs; their crystallization produces an IQH-CDW with fundamental period set by $2/\delta$, in contrast to an electron Wigner crystal whose minimum periodicity is set by $1/\delta$ (see App.~\ref{app:LSM_constraints} for a more detailed discussion).

Our numerical results open pathways for the experimental exploration of strong-coupling superconductors.
Concretely, our predictions can be explored in synthetic quantum systems such as cold atoms in optical lattices by engineering synthetic gauge fields, which allow for the highly-controllable realization of Hubbard-Hofstadter models~\cite{Aidelsburger2013,Jotzu2014,Mancini2015,Celi2014}.
In addition, layer pseudo-spin degrees of freedom, which are immune to Zeeman fields, naturally emerge in heterostructures of 2D materials~\cite{Eisenstein2014}. For sufficiently screened Coulomb interactions, certain moir\'e structures at strong magnetic fields may be approximately described by the studied model~\cite{Zhang2021,Kuhlenkamp24}.
Other multi-layer structures, such as ABA-stacked transition metal dichalcogenides or moir\'e lattices imprinted by twisted hBN~\cite{Yasuda2021,Wang2025,Kiper2025}, may be captured by Hubbard models with approximate $\mathrm{SU}(N>2)$ symmetry.

We emphasize that doping a spin liquid does not automatically lead to superconductivity, and that additional input such as the energetic hierarchy of charged anyons is crucial~\cite{DivicPNAS, Shi2024,shi2025_nonabelian}. In the studied triangular Hubbard-Hofstadter model, this hierarchy is imposed by proximity to an insulator-insulator topological critical point~\cite{DivicPNAS}. Here, we have reported a robust chiral superconductor at finite doping, with no sign of the various competing orders---such as stripe, charge-density-wave, or antiferromagnetic metals---that dominate zero-flux Hubbard systems on both the square and triangular lattices~\cite{ArovasAnnRev2022,zhu2022doped}. Instead, over the range of dopings $\delta\in \lbrace 1/15,\, 1/12,\, 1/10,\, 1/8 \rbrace$ and Hubbard strengths on both sides of the IQH-CSL transition, our results point to a unique chiral superconducting ground state. 
For the strong repulsive Hubbard strengths examined, superconductivity cannot be understood as arising from a weak-coupling instability of a Fermi surface.

The advocated criticality-assisted route should apply to other platforms and topological orders, notably doped fractional Chern insulators~\cite{Pichler2025}, and thus offers a concrete design principle for engineering robust realizations of anyon-driven superconductivity. Our model provides a clean playground for probing the fundamental interplay of quantum criticality, topological order and superconductivity, paving the way for a deeper understanding of the properties of matter arising at finite anyon density.

\bigskip
\begin{acknowledgments}
We are grateful to the authors of Ref.~\cite{Chen2025_topoSC} for sharing their manuscript with us and for coordinating submission; we reach similar conclusions.
We acknowledge helpful discussions with Sajant Anand, Marcus Bintz, Martin Claassen, Jixun K. Ding, Junkai Dong, Johannes Feldmeier, Andrew Hardy, Wilhelm Kadow, Charlie Kane, Steve Kivelson, Nandagopal Manoj, Frank Pollmann, Miles Stoudenmire, Tianle Wang, and Wyndham White.
We thank Andy Millis, Valentin Cr\'epel, Xue-Yang Song, Fabian Pichler, and Michael Knap for insightful discussions and collaborations on related topics. We further thank Andy Millis for detailed comments on the manuscript.
C.K. acknowledges funding from the Swiss National Science Foundation (Postdoc.Mobility Grant No. 217884). 
S.D. and M.P.Z. are supported by the Simons
Collaboration on Ultra-Quantum Matter, which is a
grant from the Simons Foundation (1151944, M.Z.)
A.V. is supported by a Simons Investigator award, the Simons Collaboration on Ultra-Quantum Matter, which is a grant from the Simons Foundation (651440, A.V.)
This research is funded in part by the Gordon and Betty Moore Foundation’s EPiQS Initiative, Grant GBMF8683 to T.S.
\end{acknowledgments}

\bibliographystyle{apsrev4-2} 
\bibliography{library}

\clearpage

\onecolumngrid

\section*{ Supplementary Material: ``Robust superconductivity upon doping chiral spin liquid and Chern insulators in a Hubbard-Hofstadter model''}

\setcounter{equation}{0}
\setcounter{figure}{0}
\setcounter{table}{0}

\setcounter{secnumdepth}{2}
\renewcommand{\thefigure}{S\arabic{figure}}
\renewcommand{\bibnumfmt}[1]{[S#1]}
\setcounter{section}{0}

\appendix
\restoreTOC
\tableofcontents

\clearpage

\section{Lattice and flux-threading conventions}
\label{app:lattice_conventions}

\subsection{YC geometry} \label{sec:YC_geometry}

Here we describe our conventions for the YC cylinder geometries. Beginning
in the 2D limit, we position lattice sites at the locations $\bm{r}=n_{1}\bm{a}_{1}+n_{2}\bm{a}_{2}$
where $n_{j}\in\mathbb{Z}$ and 
\begin{align}
\bm{a}_{1}=a\left(\frac{\sqrt{3}}{2}\hat{x}+\frac{1}{2}\hat{y}\right),\qquad\bm{a}_{2}=a\hat{y}\qquad\text{(YC)}.
\end{align}
The corresponding reciprocal lattice vectors are
\begin{align}
\bm{b}_{1}=\frac{4\pi}{\sqrt{3}a}\hat{x},\qquad\bm{b}_{2}=\frac{2\pi}{a}\left(-\frac{\hat{x}}{\sqrt{3}}+\hat{y}\right)\qquad\text{(YC)}.
\end{align}
In the 2D plane, we choose hoppings so that the translation symmetries are given by
\begin{align} \label{eq:YC_translations}
T_{1}c_{\bm{r}}^{\dagger}T_{1}^{\dagger}=(-1)^{n_{2}}c_{\bm{r}+\bm{a}_{1}}^{\dagger},\qquad T_{2}c_{\bm{r}}^{\dagger}T_{2}^{\dagger}=c_{\bm{r}+\bm{a}_{2}}^{\dagger}\qquad\text{(YC)}.
\end{align}

The construction of the YC-$L_{y}$ cylinder geometry consists of
``identifying'' each point $\bm{r}$ with $\bm{r}+L_{y}\bm{a}_{2},$
namely by equating the operators $c_{\bm{r}+L_{y}\bm{a}_{2}}^{\dagger}=T_{y}^{L_{y}}c_{\bm{r}}^{\dagger}T_{y}^{-L_{y}}=c_{\bm{r}}^{\dagger}$
where (to unify notation with the XC case below) we denote the circumferential
magnetic translation symmetry by $T_{y}=T_{2}.$ Since $(-1)^{n_{2}}$
is not single-valued when $L_{y}$ is odd, then $T_{1}$ is not well-defined
(let alone a symmetry) in this case, manifesting the fact that the
gauge-invariant fluxes through adjacent loops differ by $\pi$~\cite{Divic24}. However,
$T_{2}$ remains a well-defined symmetry.

Now we consider nonzero threaded flux $\Phi,$ which requires that
\begin{align} \label{eq:YC_twistbc}
    T_{y}^{L_{y}}=e^{i\Phi N_{F}},   
\end{align}
where $N_{F}$ is the fermion number operator. Demanding that $T_{y}$ continue to be defined as a pure coordinate shift, as in Eq.~\eqref{eq:YC_translations}, then the flux condition (Eq.~\ref{eq:YC_twistbc}) is equivalent to the twisted boundary condition
\begin{align} \label{eq:YC_wraparound}
c_{\bm{r}+L_{y}\bm{a}_{2}}^{\dagger}=e^{i\Phi}c_{\bm{r}}^{\dagger}\qquad\text{(YC)}.
\end{align}

\subsection{XC geometry}

Here we describe our conventions for the XC cylinder geometries. In
this case, it is convenient to orient the plane such that the lattice
sites are located at
\begin{align}
\tilde{\bm{a}}_{1}=a\hat{x},\qquad\tilde{\bm{a}}_{2}=a\left(\frac{1}{2}\hat{x}+\frac{\sqrt{3}}{2}\hat{y}\right)\qquad\text{(XC)},
\end{align}
with corresponding reciprocal lattice vectors
\begin{align}
\tilde{\bm{b}}_{1}=2\pi\hat{x}-\frac{2\pi}{\sqrt{3}}\hat{y},\qquad\tilde{\bm{b}}_{2}=\frac{4\pi}{\sqrt{3}}\hat{y}\qquad\text{(XC)}.
\end{align}
Similarly, the lattice sites are located at positions $\bm{r}=n_{1}\tilde{\bm{a}}_{1}+n_{2}\tilde{\bm{a}}_{2}.$
We choose the hoppings so that the magnetic translation symmetries
are
\begin{align}
\tilde{T}_{1}c_{\bm{r}}^{\dagger}\tilde{T}_{1}^{\dagger}=c_{\bm{r}+\tilde{\bm{a}}_{1}}^{\dagger},\qquad\tilde{T}_{2}c_{\bm{r}}^{\dagger}\tilde{T}_{2}^{\dagger}=(-1)^{n_{1}}c_{\bm{r}+\tilde{\bm{a}}_{2}}^{\dagger}\qquad\text{(XC)}.
\end{align}
(Note that $\tilde{T}_{1,2}$ here are distinct from $T'_{1,2}$ in the main text).

To construct the XC-$2L_{y}$ cylinder, we identify points separated
by the next-nearest-neighbor distance $L_{y}a\sqrt{3},$ namely $\bm{r}$
and $(2\tilde{\bm{a}}_{2}-\tilde{\bm{a}}_{1})L_{y}.$ For general
threaded flux $\Phi,$ we define the boundary conditions by
\begin{align} \label{eq:XC_twistbc}
    T_{y}^{L_{y}}=e^{i\Phi N_{F}}
\end{align}
where
\begin{align}
    T_{y}=(\tilde{T}_{2})^{2}\tilde{T}_{1}^{\dagger}   
\end{align}
is the circumferential
translation symmetry. Note that $T_{y}$ is a pure coordinate shift:
\begin{align}
T_{y}c_{\bm{r}}^{\dagger}T_{y}^{\dagger}=(\tilde{T}_{2})^{2}\tilde{T}_{1}^{\dagger}c_{\bm{r}}^{\dagger}\tilde{T}_{1}(\tilde{T}_{2})^{-2}=(\tilde{T}_{2})^{2}c_{\bm{r}-\tilde{\bm{a}}_{1}}^{\dagger}(\tilde{T}_{2})^{-2}=(-1)^{2(n_{1}-1)}c_{\bm{r}+2\tilde{\bm{a}}_{2}-\tilde{\bm{a}}_{1}}^{\dagger}=c_{\bm{r}+2\tilde{\bm{a}}_{2}-\tilde{\bm{a}}_{1}}^{\dagger}.
\end{align}
Thus, the boundary condition (Eq.~\ref{eq:XC_twistbc}) is equivalent to
\begin{align}
c_{\bm{r}+(2\tilde{\bm{a}}_{2}-\tilde{\bm{a}}_{1})L_{y}}^{\dagger}=e^{i\Phi}c_{\bm{r}}^{\dagger}\qquad\text{(XC)},
\end{align}
which is the XC analogue of Eq.~\eqref{eq:YC_wraparound} above.

\subsection{Hybrid Fourier transform}

In both the YC and XC cases, we identified a circumferential magnetic
translation operator $T_{y}$ which is a well-defined symmetry for
both even and odd $L_{y}.$ We also saw that flux threading is implemented
by demanding that
\begin{align}
T_{y}^{L_{y}}=e^{i\Phi N_{F}}.
\end{align}
Thus, flux threading modifies the eigenvalues of $T_{y}.$ In particular,
when $\Phi=\pi/2$ and $L_{y}$ is odd, then $T_{y}$ can carry eigenvalue
$-1$ in the $N_{F}=2$ (\textit{i.e.}, electron pair) sector, and likewise
when $\Phi=0$ and $L_{y}$ is even.

The circumferential symmetry $T_{y}$ can be used to define the hybrid
Fourier transform. To do this, we pass to the unified coordinate system
$(x,y)$ where $x$ is a cylinder ring index and $y$ is the circumferential
intra-ring coordinate, as in the main text. In the YC case, we simply define
\begin{align} \label{eq:xy_coords_YC}
(x,y)=(n_{1},n_{2})\qquad(\text{YC}),
\end{align}
such that $x$ is constant under $T_{2}$ and $y$ is constant under
$T_{1}.$ In the XC case, we instead choose
\begin{align} \label{eq:xy_coords_XC}
(x,y)=(2n_{1}+n_{2},-n_{1})\qquad(\text{XC)}.
\end{align}
Then $x$ is invariant under $T_{y},$ which acts as $(n_{1},n_{2})\mapsto(n_{1}-1,n_{2}+2).$
Also, the choice $y=-n_{1}$ is both incremented by 1 under $T_{y}$
and is constant under $\tilde{T}_{2}$ translations.

Having defined $(x,y)$ coordinates for each geometry, we can use
$T_{y}$ to define the Fourier transform:
\begin{align} \label{eq:hybrid_fourier}
c_{x,k_{y}}^{\dagger}=\frac{1}{\sqrt{L_{y}}}\sum_{Y=0}^{L_{y}-1}e^{-ik_{y}Y}T_{y}c_{x,y=0}^{\dagger}T_{y}^{\dagger}=\frac{1}{\sqrt{L_{y}}}\sum_{Y=0}^{L_{y}-1}e^{-ik_{y}Y}c_{x,Y}^{\dagger},
\end{align}
where we used the fact that $T_{y}$ is a pure coordinate shift. Note
that the inverse transformation is then simply
\begin{align}
c_{x,y}^{\dagger}=\frac{1}{\sqrt{L_{y}}}\sum_{k_{y}}e^{ik_{y}y}c_{x,k_{y}}^{\dagger}.
\end{align}
The allowed values of momentum $k_{y}$ are dictated by the boundary
conditions:
\begin{align}
e^{i\Phi}c_{x,y}^{\dagger}=c_{x,y+L_{y}}^{\dagger}=\frac{1}{\sqrt{L_{y}}}\sum_{k_{y}}e^{ik_{y}(y+L_{y})}c_{x,k_{y}}^{\dagger}\implies e^{i\Phi}=e^{ik_{y}L_{y}},
\end{align}
\textit{i.e.},
\begin{align}
k_{y}\in\frac{\Phi}{L_{y}}+\frac{2\pi}{L_{y}}\mathbb{Z}_{L_{y}},
\end{align}
where $k_{y}$ is ``equivalent'' to $k_{y}+2\pi$ in the sense that
$c_{x,k_{y}+2\pi}^{\dagger}=c_{x,k_{y}}^{\dagger}$ for all $\Phi,$
which follows by definition (Eq.~\ref{eq:hybrid_fourier}). Also, note that threading $\Phi=2\pi$
flux is a gauge transformation that permutes the wires.

\subsection{Particle-hole symmetry on the cylinder}

Consider the unitary particle-hole (PH) operation defined by $\tilde{\mathcal{P}} c^\dag(\bm{r}) \tilde{\mathcal{P}}^{-1} = c(\bm{r})$, with $\tilde{\mathcal{P}} i \tilde{\mathcal{P}}^{-1} = +i.$
The latter condition specifies that it is a linear (as opposed to anti-linear) unitary many-body operator. Other definitions of particle-hole symmetry may differ by a gauge transformation, which does not change the considerations below.

If 1 and 2 denote an adjacent pair of sites, then this PH operation acts as
\begin{align}
\tilde{\mathcal{\mathcal{P}}}(t_{12}c_{1}^{\dagger}c_{2}+t_{12}^{*}c_{2}^{\dagger}c_{1})\tilde{\mathcal{\mathcal{P}}}^{-1} &= -t_{12}^{*}c_{1}^{\dagger}c_{2}-t_{12}c_{2}^{\dagger}c_{1},
\end{align}
so that it reverses the Peierl's phase and shifts it by $\pi.$ On the triangular lattice, this modifies the flux through each elementary triangular plaquette as follows:
\begin{align} \label{eq:triangle_PH_action}
\Phi_{\triangle}\to-\Phi_{\triangle}+\pi.
\end{align}
Requiring that this operation leave the flux invariant modulo $2\pi,$ we find that we need
\begin{align}
2\Phi_{\triangle}=\pi\left(2n+1\right),\qquad n\in\mathbb{Z}.
\end{align}
This is consistent with the fact that the Hofstadter spectrum on the triangular lattice only has bands with opposite energy for these values of $\Phi_\triangle$~\cite{Divic24,DivicPNAS}. In particular, in this work we consider $\Phi_\triangle = \pi/2$.

On the cylinder, we must also consider the flux through every loop that winds around the finite circumference: PH symmetry requires that every even-length loop (in units of the lattice constant) contains $0$ or $\pi$ flux, while every odd-length loop contains $\pm \pi/2$ flux. For both the YC-$L_y$ and XC-$2L_y$ cylinders, PH symmetry is respected if $\phi_c = 0$ when $L_y$ is even, and if $\phi_c = \pi/2$ when $L_y$ is odd. If $L_y$ is additionally a multiple of four, then $\phi_c = \pi$ also yields a PH-symmetric cylinder. These are the values of $\phi_c$ we focus on in this study.

\section{Computational details}
\label{app:comp_details}

Our iDMRG calculations are done while explicitly conserving $\mathrm{U}_c(1)$ charge, $\mathrm{U}_{s}(1)$ spin, and $\mathbb{Z}_{L_y}$ momentum along the compact $y$ direction. All calculations are carried out using the open-source TeNPy library~\cite{tenpy,tenpy_v1}.

We focus on ground states at small doping of holes, $\delta \in [1/15,\, 1/8]$, corresponding to an average electronic density $n = 1-\delta$ per site. Motivated by the existence of a robust spin gap at half-filling, both within the Mott insulating state and near the transition~\cite{Kuhlenkamp24,Divic24,DivicPNAS}, our simulations are restricted to the $S_z=0$ sector.
On all geometries and parameter values studied, a preliminary calculation conserving only $\mathrm{U}_c(1)\times \mathrm{U}_s(1)$ indicates that the ground state is invariant under $T_y$ symmetry.
In every case, the ground state at $\delta > 0$ is obtained when the calculation is initialized with $\pi$ circumferential momentum per doped $2e$ charge relative to the parent insulator at $\delta=0$. The parent insulator has the same momentum per magnetic unit cell (\textit{i.e.}, pair of cylinder rings) as the non-interacting filled band at interaction strength $U/t=0$.

\begin{figure*}[ht]
\centering
\includegraphics[width=1.0\textwidth]{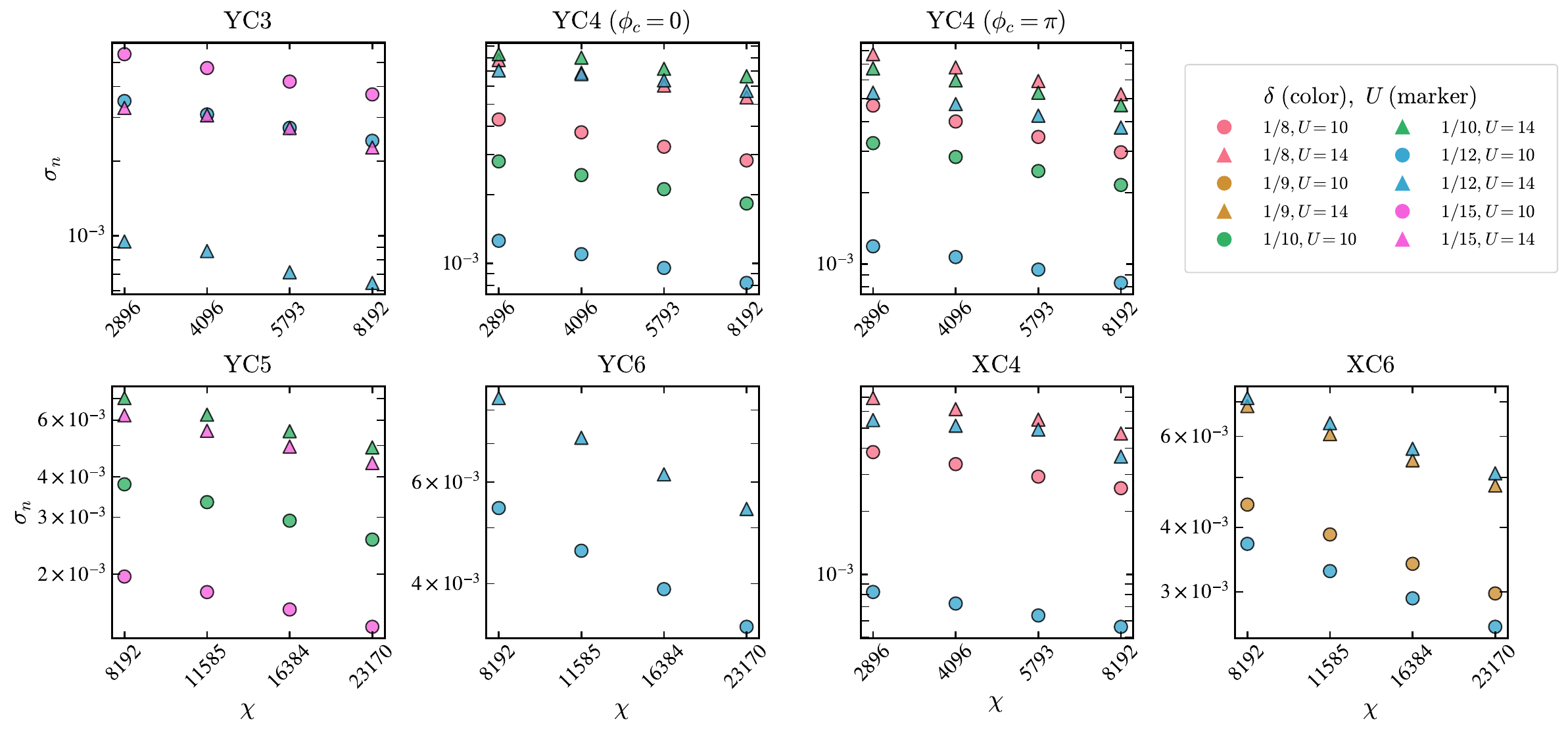}
\caption{Standard deviation $\sigma_n$ of the ring-averaged electric charge vs. MPS bond dimension for all systems (indicated in titles) and hole dopings $\delta$, at both $U/t=10$ and $14$ (indicated in legend). The decrease of $\sigma_n$ with $\chi$, which is approximately linear on the log-log plot and therefore polynomially decaying with $\chi$, indicates that the converged $\chi\to\infty$ ground state does not exhibit oscillations in the ring-averaged charge density.}
\label{fig:stddev_n}
\end{figure*}

The ground states we find have a weak translation breaking between rings at finite bond dimension. However, the symmetry breaking order parameter decays algebraically with bond dimension (Fig.~\ref{fig:stddev_n}).

\section{Precise definitions of correlation functions} \label{app:def_corr_functions}

The key diagnostic of superconducting order is the pair correlation function. Since we work in the hybrid position-momentum basis, we can readily probe the correlations of Cooper pairs $\Delta_{d}^{\dag}(x,P_y)$ carrying definite total circumferential momentum $P_y,$ labeling eigenvalues of the $T_{y}$ translation symmetry, as specified in the main text and Sec.~\ref{app:lattice_conventions} above. Here, $d$ refers to the spatial orientation of the spin-singlet Cooper pair bond being probed, as defined below.

On the YC cylinders, we evaluate pairing on bonds oriented along the $T_1$ and $T_2$ directions. In the $(x,k_y)$ hybrid coordinate parameterization of the YC cylinders, pairing along $T_1$ and $T_2$-oriented bonds correspond to
\begin{equation}
\begin{aligned}
\Delta_{1}^{\dag}(x,P_y) & = L_{y}^{-1}\sum_{k_y}c^{\dagger}(x,P_y-k_y)\, is_y c^{\dag}(x+1,k_y), \\
\Delta_{2}^{\dag}(x,P_y) & = L_{y}^{-1}\sum_{k_y} e^{ik_y} c^{\dag}(x,P_y-k_y)\, is_{y} c^{\dag}(x,k_y),
\end{aligned}
\end{equation}
which we call pairing in the $d=1$ and $d=2$ directions, respectively.
Here $s_y$ denotes the $y$ Pauli matrix for spin, $c = (c_\uparrow,c_\downarrow)$, and summation over spin indices is implicit. For the XC cylinders, we similarly consider pairing along $T_1'$ (\textit{i.e.}, in the $d=1'$ direction):
\begin{equation}
\Delta_{1'}^{\dag}(x,P_y)=L_{y}^{-1}\sum_{k_y}c^\dag(x,P_y-k_y)\, is_{y} c^\dag(x+1,k_y).
\end{equation}
We will study the ground state pair correlation functions
\begin{align}
C^{P_y,d}_{\Delta}(x;x_0) = \langle\Delta_{d}(x_0,P_y)\Delta_{d}^{\dag}(x_0 + x,P_y)\rangle,
\end{align}
where $x\in\mathbb{Z}$ now represents the separation between the electron and hole pairs, in units of cylinder rings.

We will also analyze the correlations of the ring-averaged charge and spin densities, namely
\begin{align}
    n(x) &= L_y^{-1} \sum_{y} \sum_{\sigma, y} c^\dagger_\sigma(x,y) c_\sigma(x,y), \\ 
    S_z(x) &= L_y^{-1} \sum_{y} c^\dagger(x,y) \frac{s_z}{2} c(x,y)
\end{align}
where $s_z$ is the $z$ Pauli matrix for spin. We denote the connected density correlations by
\begin{align}
    C_n(x;x_0) = \langle n(x_0) n(x_0+x)\rangle - \langle n(x_0) \rangle\langle n(x_0+x)\rangle,   
\end{align}
and likewise for $S_z$. Finally, we will analyze the electron-hole correlations
\begin{align}
    C_{c^\dag}(x;x_0) = L_y^{-2} \sum_{k_y, \sigma} |\langle c^\dag_\sigma(x_0,k_y) c_\sigma(x_0+x,k_y) \rangle |.
\end{align}

Due to weak translation symmetry breaking at finite MPS bond dimension $\chi$, the correlation functions $C_\mathcal{O}(x;x_0)$ ($\mathcal{O}=\Delta,n,S_z,c^\dag$) depend on the initial ring $x_0$.
We thus perform an average over $x_0$. When $L_y$ is even, then $T_1$ ($T_1'$) is a symmetry of the YC (XC) cylinder, so we average over all $x_0$ in the MPS unit cell~\cite{KarraschMoore2012}. When $L_y$ is odd, so that the cylinder compactification explicitly breaks the single-ring translation symmetry while respecting its square~\cite{Divic24}, we instead average $C_\mathcal{O}(x;x_0)$ over all \textit{even} starting rings. We denote the averaged correlation functions by $C_\mathcal{O}(x)$.

\section{Extracting spin quantum Hall response}
\label{app:spin_response}

We extract the spin quantum Hall conductance of the superconducting states by computing the amount of spin that is pumped by an adiabatic insertion process.
The superconducting states we find are expected to have a unique ground state and no fractionalized excitations. They should feature a non-fractionalized quantized spin quantum Hall response, according to Laughlin's pumping argument~\cite{laughlin1981quantized}. For the same reason, the Hamiltonian on the cylinder should have a unique, albeit gapless, ground state. The adiabatic flux threading process can then be emulated by finding the ground state at each $\Phi_s$. The amount of pumped spin $\Delta S_z$ can be computed from the MPS as
\begin{equation}
    \Delta S_z(\Phi_s) = \sum_j S_{z,j}|\lambda_j(\Phi_s)|^2,
\end{equation}
where $j$ labels the Schmidt states at a particular spatial partition of the cylinder into two semi-infinite halves, $\lambda_j$ are the associated singular values, and $S_{z,j}$ label the spin quantum numbers of the Schmidt states~\cite{Zaletel2014}. The results, shown in the main text (Fig.~\ref{fig:spin_pump}), reveal a spin quantum Hall conductance $\sigma^s_{xy} = 2\times (\hbar/2)^2 / h$, consistent with the prediction of Ref.~\cite{DivicPNAS} obtained by examining the lowest-lying Cooper pair excitations of the insulating states.

\section{Proxy for Pair wavefunction}
\label{app:pair_function}

\begin{figure*}[ht]
\centering
\includegraphics[width=.5\textwidth]{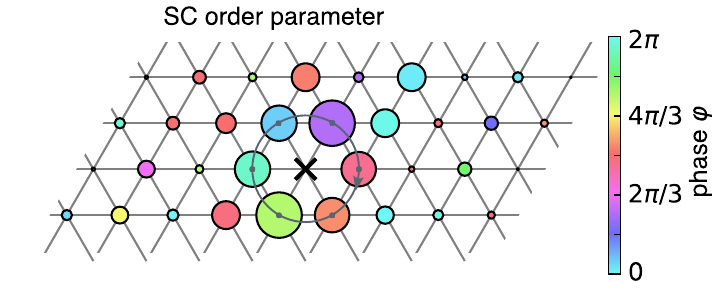}
\caption{\textbf{Proxy for the pair wavefunction in the IQH limit.} Same as right panel of Fig.1 in the main text, but evaluated for $U/t=10$ at $\chi=6144$.}
\label{fig_app:pair_wavefunction}
\end{figure*}

Here we present details on the evaluated proxy for the the 2D order parameter
$\Delta_{\v{0}}(\delta\mathbf{r}) =  \langle c(\delta\mathbf{r})\, is^y \, c(\v{0}) \rangle_\mathrm{2D}$.
We work in the $C_{6z}$ symmetric gauge of Ref.~\cite{DivicPNAS}; the $C_2$ eigenvalue of the pair correlation function was shown to be independent of the choice of gauge.
Following Ref.~\cite{Sahay23}, we consider cylinder correlation functions of the form $\langle \Delta^\dagger_{\mathbf{r}_0}(\mathbf{a})\, \Delta_{\mathbf{0}}(\delta\mathbf{r})\rangle,$
where $\Delta^\dagger_{\mathbf{r}_0}(\mathbf{a})$ is a reference Cooper pair located a distance $\mathbf{r}_0$ from the origin. This captures the off-diagonal quasi-long-range order of the system. In the limit $L_y \gg 1$ and $|\mathbf{r}_0| \gg \xi_P$, where $\xi_P$ is the spatial extent of the Cooper pair,
we expect 
\begin{equation}
 \langle \Delta^\dagger_{\mathbf{r}_0}(\mathbf{a})\, \Delta_\mathbf{0}(\delta\mathbf{r})\rangle \approx B_{\mathbf{r}_0,\mathbf{a}} \cdot\Delta_\mathbf{0}(\delta\mathbf{r}) +\mathcal{O}(e^{-|\mathbf{r_0}|/\xi_P}),
\end{equation}
up to a non-universal constant $B_{\mathbf{r}_0,\mathbf{a}}$~\cite{Sahay23}. 
We evaluate the left side of this expression
on the YC4 cylinder at $\delta=1/8$ hole doping of the CSL phase at $U/t=14$, in a $C_{6z}$ invariant gauge~\cite{DivicPNAS}, which is plotted in Fig.~\ref{fig:mainmessage}(b). This reveals that the superconducting pair wavefunction is odd under spatial inversion. Even though $C_{3z}$ symmetry is broken by the cylinder compactification, the pair correlation function winds by $2\pi$ as $\delta\mathbf{r}$ traverses, clockwise, the six nearest neighbors of the origin. This highlights the chiral $p$-wave nature of the Cooper pair, previously predicted in Ref.~\cite{DivicPNAS}. A similar calculation can be done on the IQH side of the transition at $U/t=10$, which is shown in Fig.~\ref{fig_app:pair_wavefunction}, and yields the same qualitative behaviour, consistent with the expectation that both superconductors are the same. We remark that the precise phases depend on the gauge choice of the background magnetic flux in the Hubbard model.

\section{Additional Correlation Data}
\label{eq:app_correlation_data}

In this section, we present additional data on correlation functions and correlation length for a range of dopings for different cylinder sizes. In all the systems we study, we find $\Delta$ with $k_y = \pi$ is always the dominant one.

\begin{figure*}[ht]
\centering
\includegraphics[width=1.0\textwidth]{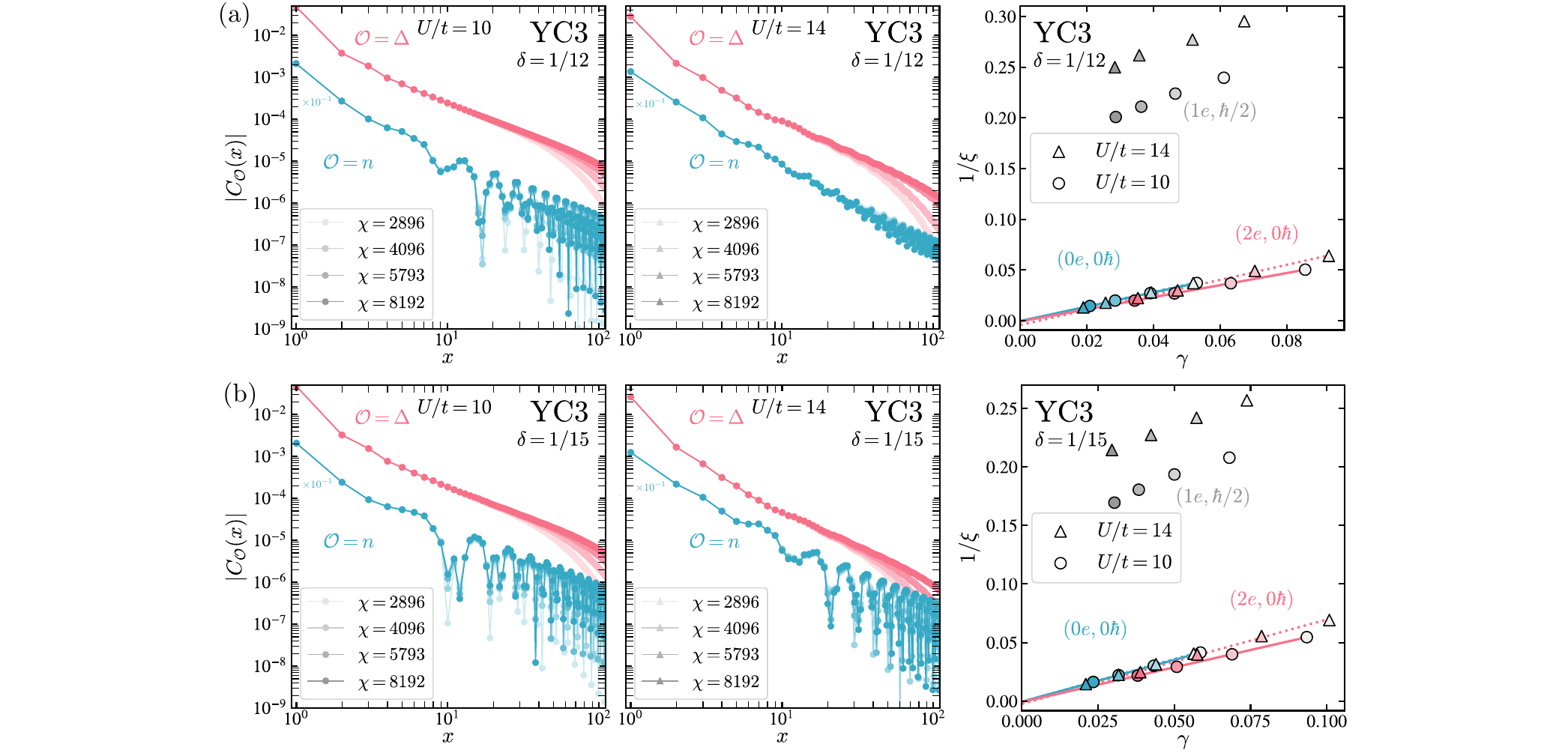}
\caption{\textbf{Pair and density correlations on the YC3 cylinder.} (a) Left panel: Magnitude of pair correlation function (red) and connected ring-averaged charge density correlations (blue) as a function of separation (in units of rings) along the YC3 cylinder at $\delta=1/12$ hole doping, Hubbard strength $U/t=10$, and MPS bond dimensions $\chi$ indicated in legend (increasing light to dark). Density correlations scaled by $10^{-1}$ for ease of viewing.
Middle panel: Same at $U/t=14$.
Right panel: inverse of dominant correlation length vs. refinement parameter $\gamma$ (see main text and Ref.~\cite{rams2018precise,Vanhecke2019scaling}) for same MPS bond dimensions as left panel, in various transfer matrix charge sectors (labeled), for $U/t=14$ (triangles) and $10$ (circles). Linear fits in the $(0e,0\hbar)$ and $(2e,0\hbar)$ sectors, at both $U/t=14$ (dotted lines) and $10$ (solid lines).
(b) Same at hole doping $\delta = 1/15$.
In all cases, pairing correlations are evaluated for $\Delta_d(x,k_y)$ with direction $d=2$ and dominant pairing momentum $k_y=\pi$.
}
\label{fig:corrs_xis_YC3_SM}
\end{figure*}

\begin{figure*}[ht]
\centering
\includegraphics[width=1.0\textwidth]{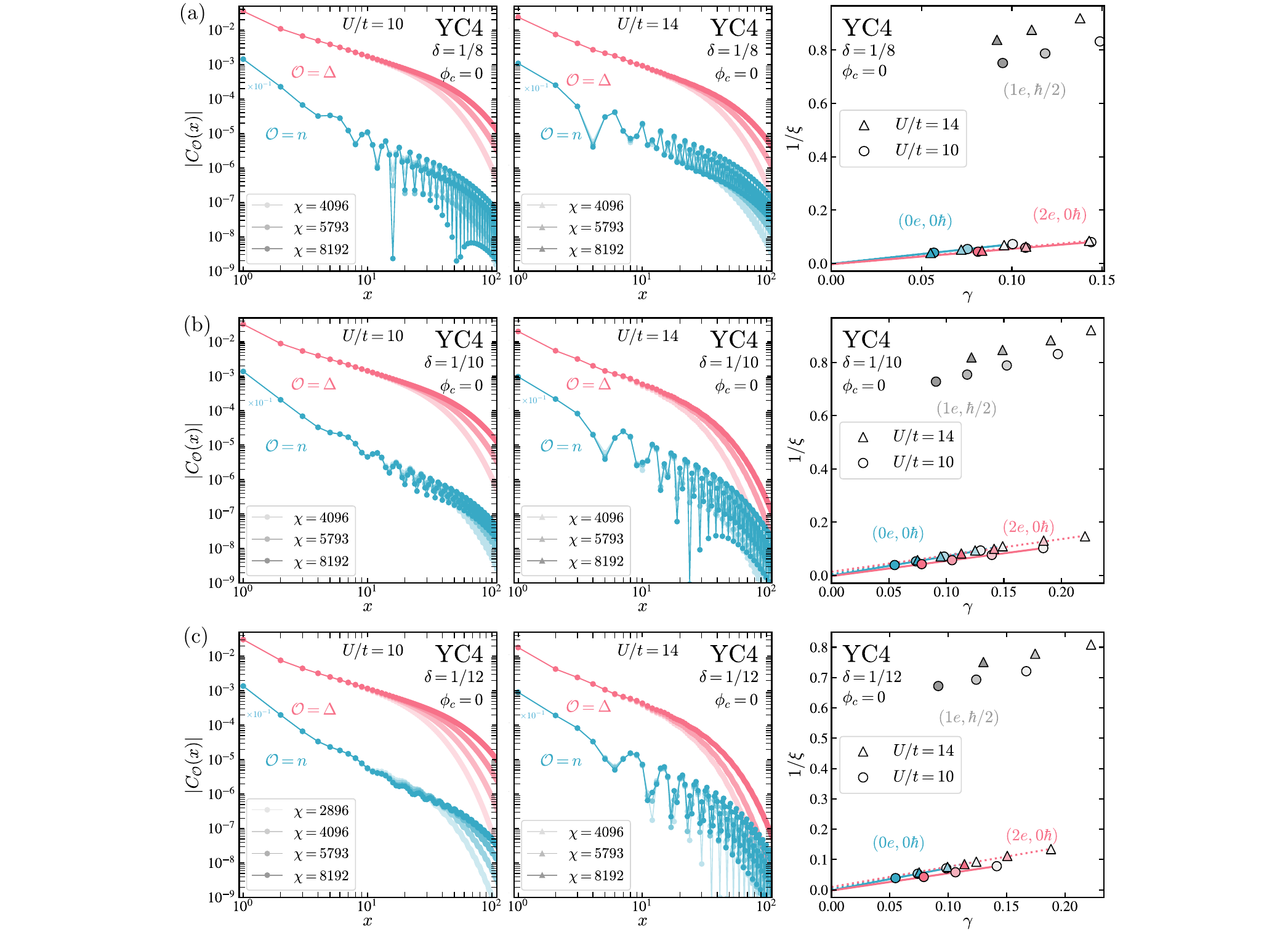}
\caption{Same as Fig.~\ref{fig:corrs_xis_YC3_SM} but for the infinite YC4 cylinder with $\phi_c=0$ at the indicated $\delta$ and $U$ values. In all cases, pairing correlations are evaluated for $\Delta_d(x,k_y)$ with direction $d=2$ and dominant pairing momentum $k_y=\pi$.}
\label{fig:corrs_xis_YC4_phi=0_SM}
\end{figure*}

\begin{figure*}[ht]
\centering
\includegraphics[width=1.0\textwidth]{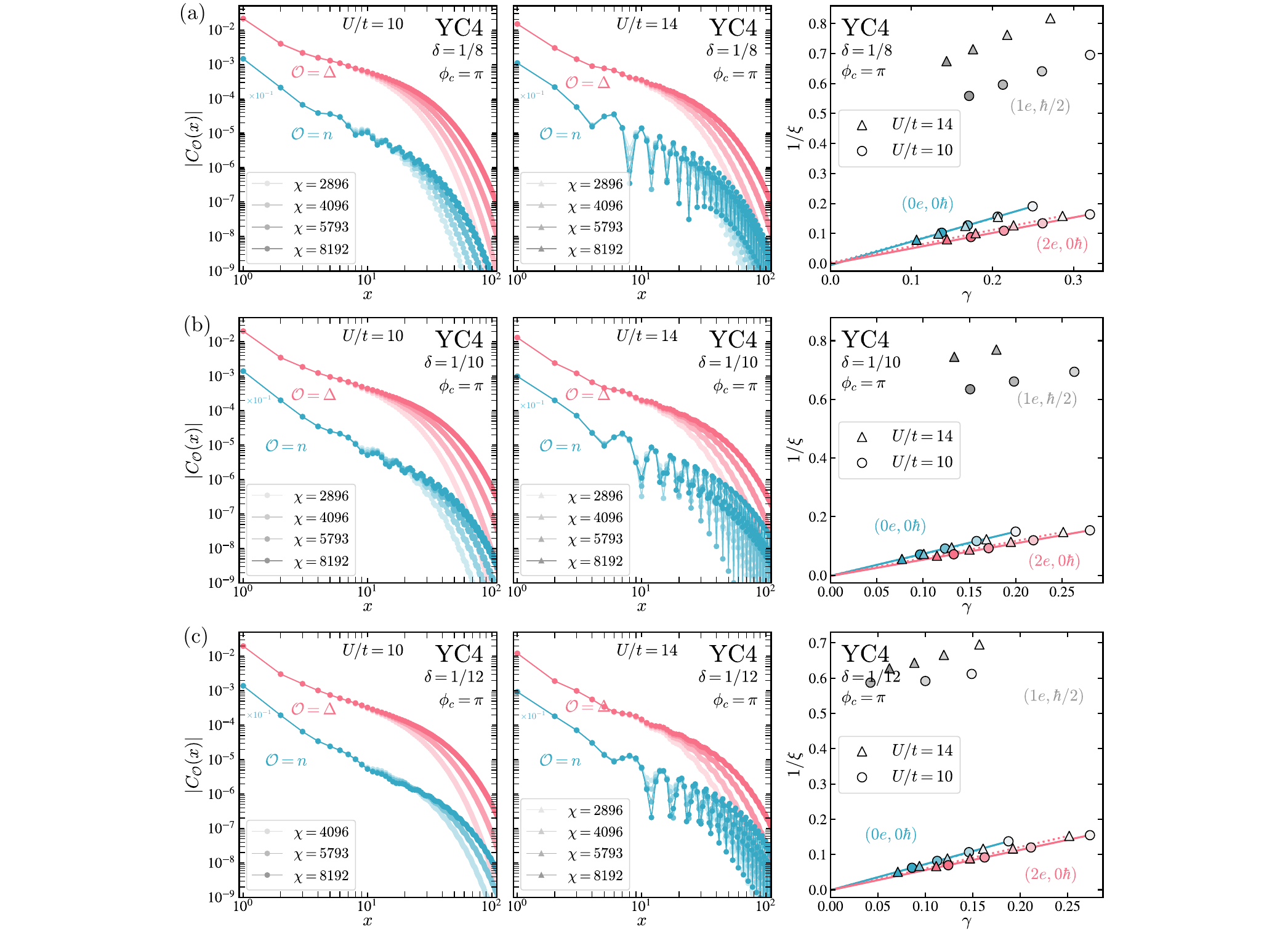}
\caption{Same as Fig.~\ref{fig:corrs_xis_YC3_SM} but for the infinite YC4 cylinder with $\phi_c=\pi$ at the indicated $\delta$ and $U$ values. In all cases, pairing correlations are evaluated for $\Delta_d(x,k_y)$ with direction $d=2$ and dominant pairing momentum $k_y=\pi$.}
\label{fig:corrs_xis_YC4_phi=pi_SM}
\end{figure*}

\begin{figure*}[ht]
\centering
\includegraphics[width=1.0\textwidth]{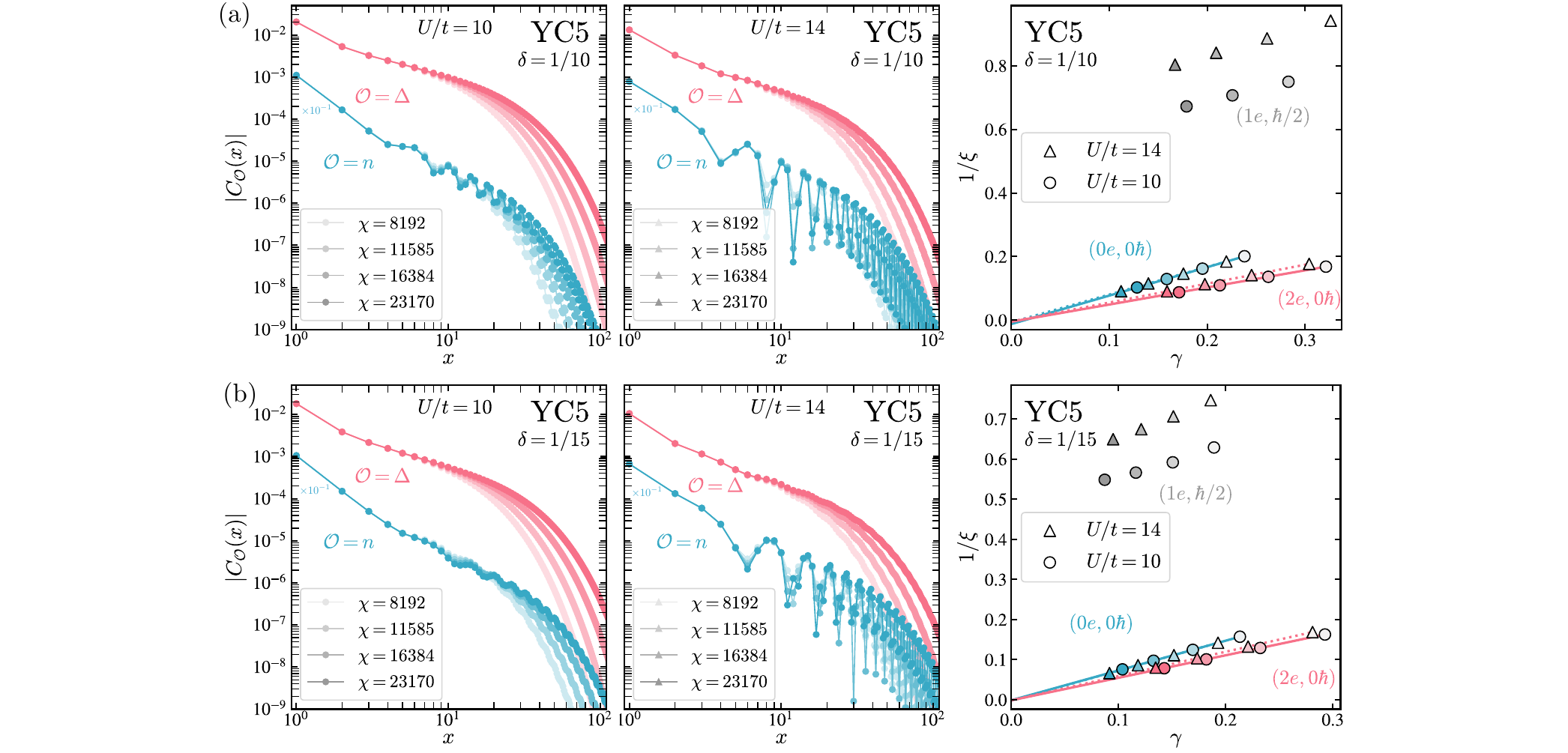}
\caption{Same as Fig.~\ref{fig:corrs_xis_YC3_SM} but for the infinite YC5 cylinder with $\phi_c=\pi/2$ at the indicated $\delta$ and $U$ values. In all cases, pairing correlations are evaluated for $\Delta_d(x,k_y)$ with direction $d=2$ and dominant pairing momentum $k_y=\pi$.}
\label{fig:corrs_xis_YC5_SM}
\end{figure*}

\begin{figure*}[ht]
\centering
\includegraphics[width=1.0\textwidth]{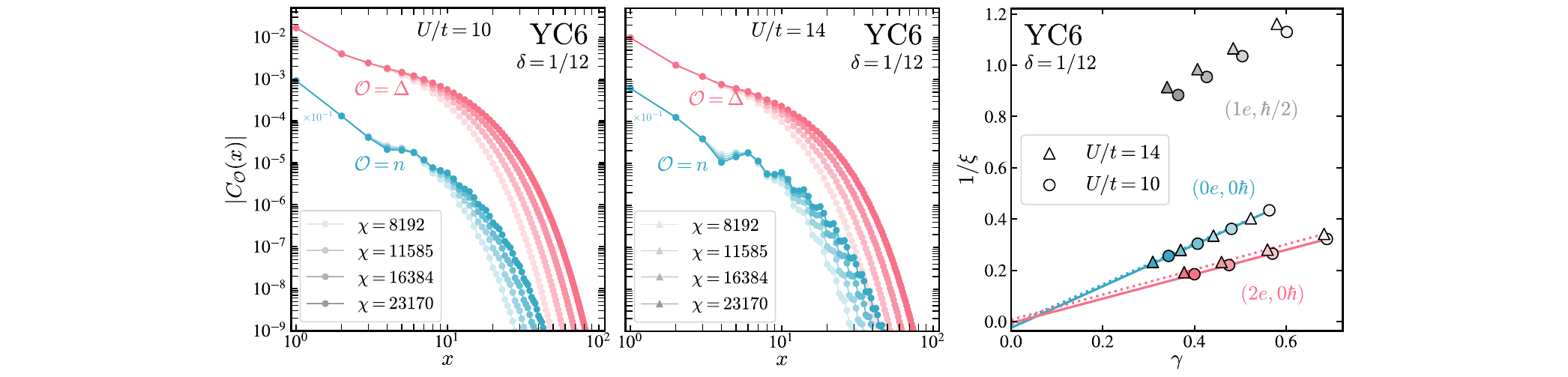}
\caption{Same as Fig.~\ref{fig:corrs_xis_YC3_SM} but for the infinite YC6 cylinder with $\phi_c=0$ at the indicated $\delta$ and $U$ values. In both cases, pairing correlations are evaluated for $\Delta_d(x,k_y)$ with direction $d=2$ and dominant pairing momentum $k_y=\pi$.}
\label{fig:corrs_xis_YC6_SM}
\end{figure*}

\begin{figure*}[ht]
\centering
\includegraphics[width=1.0\textwidth]{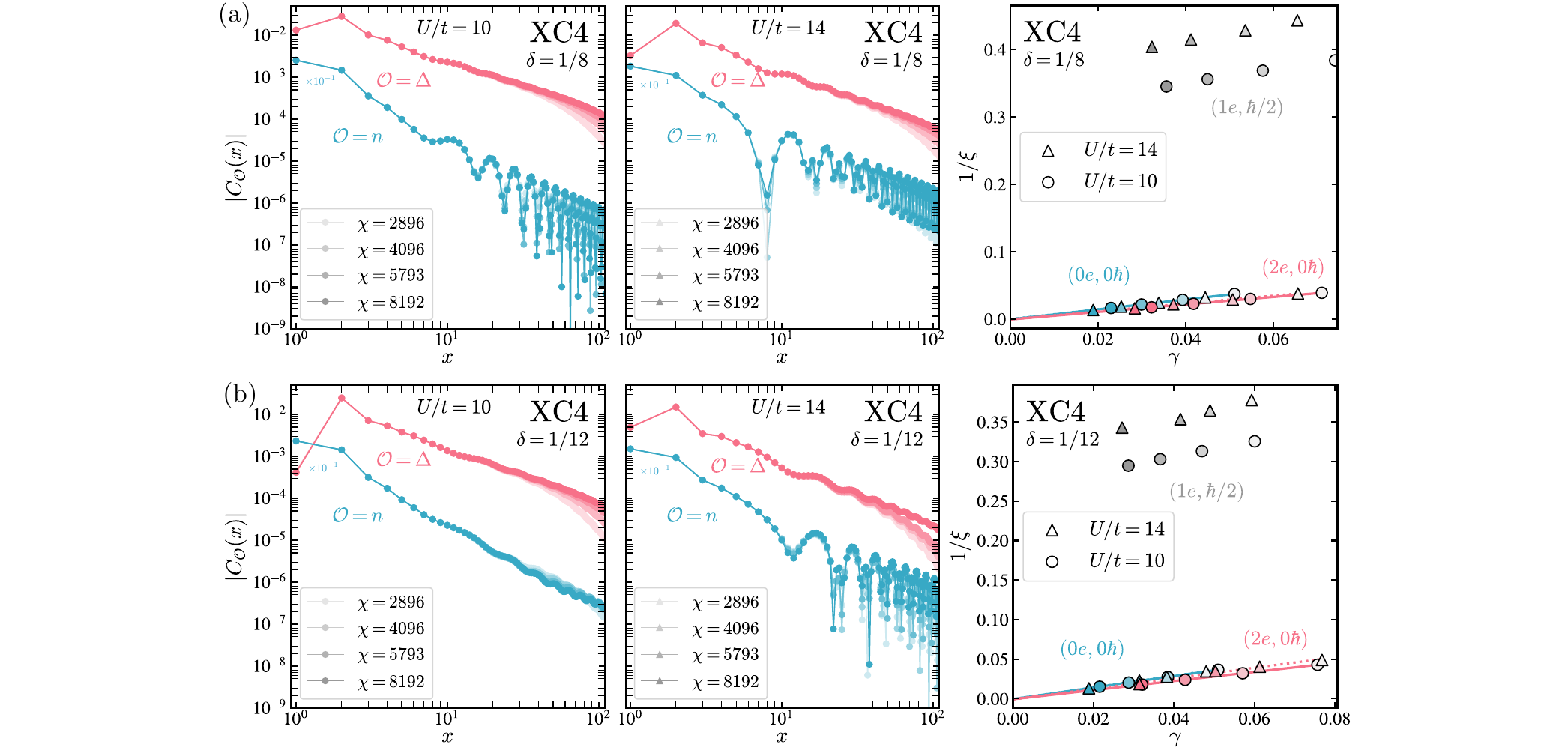}
\caption{Same as Fig.~\ref{fig:corrs_xis_YC3_SM} but for the infinite XC4 cylinder with $\phi_c=0$ at the indicated $\delta$ and $U$ values. In all cases, pairing correlations are evaluated for $\Delta_d(x,k_y)$ with direction $d=1'$ and dominant pairing momentum $k_y=\pi$.}
\label{fig:corrs_xis_XC4_SM}
\end{figure*}

\begin{figure*}[ht]
\centering
\includegraphics[width=1.0\textwidth]{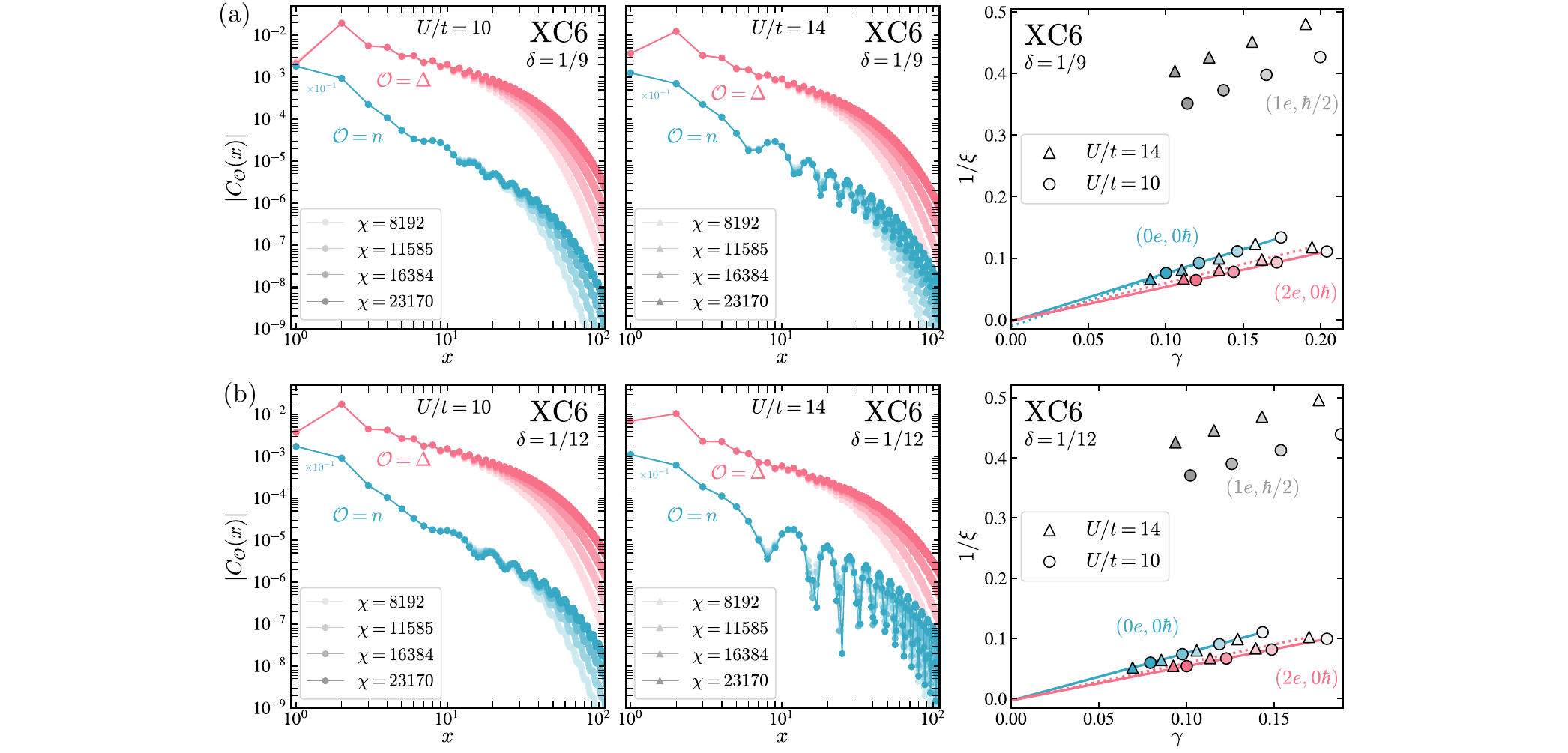}
\caption{Same as Fig.~\ref{fig:corrs_xis_YC3_SM} but for the infinite XC6 cylinder with $\phi_c=\pi/2$ at the indicated $\delta$ and $U$ values. In all cases, pairing correlations are evaluated for $\Delta_d(x,k_y)$ with direction $d=1'$ and dominant pairing momentum $k_y=\pi$.}
\label{fig:corrs_xis_XC6_SM}
\end{figure*}

\clearpage

\section{Details on extracting superconducting exponent}
\label{app:fitting_procedure}

Here we review details on extracting the power law exponents $K_\mathrm{SC}$, by scaling analysis. As we approach the critical point by increasing bond dimension $\chi$ and therby also $\xi_{2e}$, one can postulate an ansatz
\begin{equation}
    C_{\Delta}(x) = \xi_{2e}^{-K_\mathrm{SC}} F(x/\xi_{2e}),
\end{equation}
with a scaling function $F(x/\xi_{2e})$. Operationally we follow the analysis presented in~\cite{Sahay23}. For a given bond-dimension $\chi$, we extract the correlation length $\xi_{2e}$ from the transfer matrix spectrum and determine $C_\Delta^{(\chi)}(x/\xi_{2e})$.  Since originally $x\in \mathbb{Z}$, we interpolate this data linearly on a logarithmic scale such that we can compare the scaling function $F(x/\xi_{2e})$ that we obtain for different bond dimensions. To find the right power law exponent $K_\mathrm{SC}$, we numerically minimize the collapse error
\begin{equation}
    E(K) = \sum_{ \chi\neq \chi'}\int_{y_0}^{y_1} \mathrm{d}y \frac{|F^{(\chi)}(y)-F^{(\chi')}(y)|^2}{\xi_\chi\xi_{\chi'}} = \sum_{ \chi\neq \chi'}\int_{y_0}^{y_1} \mathrm{d}y \frac{|\xi_{\chi}^{K}\tilde{C}^{(\chi)}(y)-\xi_{\chi'}^{K}\tilde{C}^{(\chi')}(y)|^2}{\xi_\chi\xi_{\chi'}},
\end{equation}
which also allows us to estimate the error in our fit as
\begin{equation}
    \sigma_{K_{\mathrm{SC}}}  =\sqrt{ \frac{E(K_{\mathrm{SC}})} {\partial_K^2E(K)|_{K_{SC}}} }.
\end{equation}

In the rest of this section we present the fits yielding our estimates for $K_\mathrm{SC}$ presented in the main text.

\begin{table}[ht]
\centering
\begin{tabular}{|c|c|c|c|c||c|c|}\multicolumn{7}{c}{$K_\mathrm{SC}$ upon doping the IQH state at $U=10t$} \\ 
\hline
doping $\delta$ & YC3 & YC4 & YC5 & YC6 & XC4 & XC6\\
\hline
$1/15$ & $(1.36\pm 0.04)^\dagger$& - & $(0.97\pm0.03)^*$ & -  & - & - \\
$1/12$ & $(1.35\pm0.02)^\dagger$ & $(1.09\pm0.01)^\dagger$ &  -& $(0.68\pm 0.08)^*$ & $(1.21\pm0.03)$ & $(1.01\pm 0.04)$ \\
$1/10$  & - & $(1.08\pm0.01)^\dagger$ & $(0.91\pm0.02)^*$ & - & - & - \\
\hline
\end{tabular}
\caption{Entries are given by $K_\mathrm{SC}$ where * indicates values obtained in pairing direction `$1$', while $\dagger$ denotes pairing direction `$2$'. Pair-correlations in the XC geometries are evaluated along the `$1'$' direction.}
\label{tab:Ksc10}
\end{table}

\begin{table}[ht]
\centering
\begin{tabular}{|c|c|c|c|c||c|c|}\multicolumn{7}{c}{$K_\mathrm{SC}$ upon doping the CSL state at $U=14t$} \\ 
\hline
doping $\delta$ & YC3 & YC4 & YC5 & YC6 & XC4 & XC6\\
\hline
$1/15$ & $(1.60\pm0.04)^\dagger$& - &  $(1.31\pm0.03)^\dagger$& -  & - & - \\
$1/12$ & $(1.66\pm0.04)^\dagger$ & $(1.33\pm 0.02)^\dagger$&  -& $(0.79\pm0.07)^*$ & $(1.42\pm0.04)$ & $(1.11\pm0.02)$ \\
$1/10$  & - & $(1.27\pm0.01)^\dagger$ &$(1.13\pm0.03)^\dagger$  & - & - & - \\
\hline
\end{tabular}
\caption{Entries are given by $K_\mathrm{SC}$ where * indicates values obtained in pairing direction `$1$', while $\dagger$ denotes pairing direction `$2$'. Pair-correlations in the XC geometries are evaluated along the `$1'$' direction.}
\label{tab:Ksc14}
\end{table}

\begin{figure*}[!h]
\centering
\includegraphics[width=0.99\textwidth]{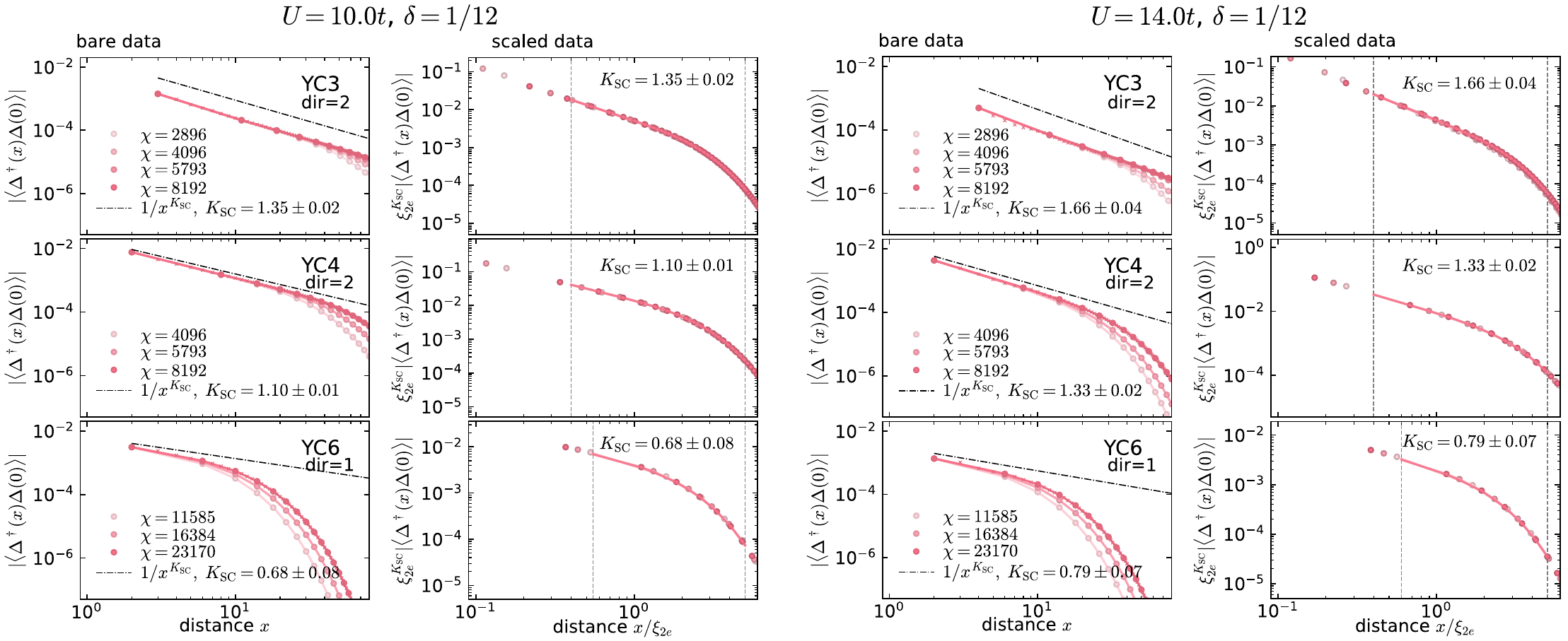}
\caption{\textbf{Detailed fitting procedures for SC correlations on the YC geometry} Power-law extraction of the superconducting correlations for $\delta=1/12$ at $U/t=10$ (left) and $U/t=14$ (right). We show superconducting correlations and their scaling collapse on YC3, YC4 and YC6 from top to bottom. The power law obtained by the optimal scaling collapse is plotted as a black dashed line along with the bare data. Due to oscillations of the correlations with period $1/\rho_\text{pair}^{1D}$, we sample points separated by this period, chosen so that the maxima are captured. All data is shown in addition as faint crosses for the highest respective MPS bond dimension $\chi$. The pairing correlations are evaluated for $\Delta_d(x,k_y)$ with direction $d=$dir (indicated in each plot) and dominant pairing momentum $k_y=\pi$.}
\label{fig:fitting_SC}
\end{figure*}

\newpage

\section{Lieb-Shultz-Mattis-type constraints for IQH-CDW and CSL-CDW}
\label{app:LSM_constraints}
Let us first define symmetry breaking patterns in the presence of uniform magnetic field. Let $\{\v{a}_i\}$ denote the set of Bravais lattice vectors for the triangular lattice. They satisfy magnetic translation algebra

\begin{equation}
    T_{\v{a}_i} T_{\v{a}_j} = e^{i\frac{\pi}{A_{uc}}\v{a}_i \wedge \v{a}_j} T_{\v{a}_j} T_{\v{a}_i},
\end{equation}
where $A_{uc}$ is the area of the parallelogram formed by two triangular plaquettes, and $\v{a}_i \wedge \v{a}_j$ is the signed area of the parallelogram formed by $\v{a}_i, \v{a}_j$.

To characterize the symmetry of the ground states, we denote by $\{\v{b}_j\} \in \{\v{a}_j\}$ the set of Bravais lattice vectors that satisfy
\begin{equation}
    T_{\v{b}_j} \ket{\Psi} \propto \ket{\Psi},
\end{equation}
where $\ket{\Psi}$ is a ground state of the many-body Hamiltonian. For undoped IQH and CSL, we have $\{\v{b}_j\} = \{\v{a}_j\}$.

This allows us to define two different notions of unit cell size. The former is the size of the primitive unit cell of the lattice formed by $\{ \v{b}_j \}$. This is the periodicity of any physical observable, since their expectation value is unchanged by the action of $T_{\v{b}_j}$ on $\ket{\Psi}$. We denote the size of this unit cell by $N_{\mathrm{physical}}$ i.e. $N_{\mathrm{physical}} A_{uc} = |\v{b}_1 \wedge \v{b}_2|$. The IQH and CSL phases have $N_{\mathrm{physical}} = 1$.

On the other hand, for the purpose of computing filling constraints, the relevant unit cell is formed by \textit{commuting} magnetic translations. To find this, find the minimum parallelogram in the lattice formed by $\{\v{b}_j\}$ such that they commute. Using $\v{c}_1, \v{c}_2$ denote the sides of this (non-unique) parallelogram, we define the size of the magnetic unit cell by $N_{\mathrm{mag}} A_{uc} = |\v{c}_1 \wedge \v{c}_2|$. From the magnetic translation algebra, we note that $N_{\mathrm{mag}} \in 2\mathbb{Z}$. The IQH and CSL phases have $N_{\mathrm{mag}} = 2$.

There is a simple relationship between $N_{\mathrm{mag}}$ and value of $N_{\mathrm{physical}}$. If $N_{\mathrm{physical}} \in 2\mathbb{Z}$, then $N_{\mathrm{physical}} = N_{\mathrm{mag}}$. Otherwise $2N_{\mathrm{physical}} = N_{\mathrm{mag}}$. Inverting this, we have
\begin{equation}
    N_{\mathrm{physical}}= \begin{cases}
         {N_{\mathrm{mag}}} &\text{if } N_{\mathrm{mag}} \in 4\mathbb{Z}
         \\
          \frac{N_{\mathrm{mag}}}{2} &\text{if } N_{\mathrm{mag}} \notin 4\mathbb{Z}
    \end{cases}
\end{equation}

Let us now consider filling constraint for general values of doping. Let $1-\delta = \frac{p}{q}, \quad p, q \in \mathbb{Z}$ denote the number of electrons per original plaquette. The filling constraint for spin-singlet topologically trivial states are given by
\begin{equation}
    N_{\mathrm{mag}} (1 - \delta) = 0 \mod 2 \implies N_{\mathrm{mag}} \delta = 0 \mod 2 \implies \min N_{\mathrm{mag}} = 2q
\end{equation}
where the first implication follows from $N_{\mathrm{mag}} \in 2\mathbb{Z}$. Denoting $\delta = \frac{p}{q}$, the minimum possible periodicity of the physical observables are given by
\begin{equation}
    \min N_{\mathrm{physical}} =
    \begin{cases}
        2q & \text{    if    } q \in 2\mathbb{Z},
        \\
        q & \text{    if    } q \notin 2\mathbb{Z},
    \end{cases}
\end{equation}

On the other hand, in the CSL phase, the existence of the Mott limit allows us to directly consider a constraint on $N_{\mathrm{physical}}$. This yields the following:
\begin{equation}
    N_{\mathrm{physical}} (1 - \delta) = 0 \mod 1 \implies N_{\mathrm{physical}} \delta = 0 \mod 1 \implies \min N_{\mathrm{physical}} = q.
\end{equation}
We see that the constraint for IQH-CDW and CSL-CDW differs \textit{iff} $q \in 2\mathbb{Z}$.

\section{Topological criticality and relation to anyon superconductivity}
\label{app:criticality_and_anyonSC}

Here we review Laughlins `anyon superconductivity' and highlight its connection to the IQH to CSL phase transition~\cite{LaughlinRelationshipHighTemperatureSuperconductivity1988,LaughlinSuperconductingGroundState1988,Song2021,DivicPNAS,Pichler2025}.

\subsection{Field theory for IQH-CSL transition} \label{app:field_theory}

In anticipation of the fractionalization in the CSL, we construct a parton ansatz in which the electron operator $c_{i,\sigma}$ is decomposed as $ c_{i,\sigma} = f_{i\sigma} b_i$, where $f_{i,\sigma}$ is a neutral fermionic spinon and $b$ is a bosonic rotor degree of freedom, which carries the charge of the electron. The redundant description is accompanied by an emergent $\mathrm{U(1)}$ gauge field $a$. This implements a gauge constraint enforcing the spinon and chargon densities to be equal equal $\sum_\sigma f_\sigma^\dagger(x)f_\sigma(x) = n_b(x)$. Throughout the phase transition, the spinons $f$ fill topological bands with Chern number $C_\uparrow = C_\downarrow =1$, inherited from the electronic band structure, while the chargon $b$ undergoes a superfluid to Mott transition~\cite{KuhlenkampThesis,DivicPNAS}. Indeed, if the chargon is condensed the gauge field $a$ is confined and the spinon can be identified as the electron. However, if $b$ is in a Mott state, this also constrains the spinon-density on each site by virtue of the constraint, leading to the formation of a CSL~\cite{WenChiralSpinStates1989}. Since the spinons remain gapped throughout the transition they can be integrated out, yielding an effective Lagrangian
\begin{equation}
\begin{aligned}
\mathcal{L}_{\mathrm{QCP}}[A,A_s] &=  |(\partial_\mu - iA_\mu  + i a_\mu ) \varphi |^2 + r |\varphi|^2 - \lambda |\varphi|^4 \\
    & - \frac{1}{4\pi} \sum_{\sigma\in \uparrow,\downarrow} \alpha_\sigma \wedge \mathrm{d}\alpha_\sigma +\frac{1}{2\pi}a\wedge\mathrm{d}(\alpha_\uparrow+\alpha_\downarrow) + \frac{1}{2\pi}A_s\wedge\mathrm{d}(\alpha_\uparrow-\alpha_\downarrow)+\dots
    \label{eq:effective_action_CSL}
\end{aligned}
\end{equation}
where $\varphi$ is a complex scalar describing the Higgs transition of the chargon, $r, \lambda$ are effective parameters driving the transition, $A$ is the external gauge field that couples to charge, $A_s$ is the external gauge field that couples to spin, and $a_\mu$ is the emergent $U(1)$ gauge field. As discussed, in the Higgs phase when $\langle \varphi \rangle \neq 0$ Eq.~\eqref{eq:effective_action_CSL} the Meissner response sets $a =A$ and recovers the K-matrix of the spin-singlet IQH state. On the other hand, when we enter the Mott phase, $\langle \varphi \rangle = 0 $ which allows $a$ to fluctuate, forcing the system into the CSL semion topological order. Moreover, Eq.~\eqref{eq:effective_action_CSL} encodes the fact that $Q=2e$ binds to unit flux of $a$, allowing for the identification of the Cooper pair excitation on both sides of the transition within the low-energy theory~\cite{Lee2018,DivicPNAS}. 
While the two phases differ in chiral central charge ($c_- = 2$ for the IQH and $c_- = 1$ for the CSL) they exhibit the same spin quantum Hall response of $\sigma_{xy}^s=2\cdot (\hbar/2)^2/h$. A continuous transition between the two phases in the Hubbard-Hofstadter model is consistent with current numerics~\cite{Kuhlenkamp24,Divic24}, while evidence for the existence of continuous IQH-CSL transitions was recently presented in Ref.~\cite{Zhou_deconf_2025}. 

\subsection{Anyon superconductivity upon doping the CSL}
When doping the system close to the critical point such that $n_e  = 1-\delta$ the constraint implies that both the spinons and chargon density must change accordingly. Even at finite doping the spinons $f_\sigma$ can remain gapped if the internal magnetic field is adjusted by $(\nabla\times a)/2\pi = -\delta/2 $, to lower the degeneracy of their Chern bands. While this perserves the spin-gap, the change in internal magnetic field is seen by the chargons $\varphi$, which see the opposite flux corresponding to finite filling $\nu_\varphi = -2$. As such they can naturally form a bosonic integer quantum Hall state, which when glued back to the spinon-theory describes a supercondcutor~\cite{Song2021}
\begin{equation}
\begin{aligned}
\mathcal{L}^\text{CSL}_{\mathrm{doped}}[A,A_s] &=  \frac{-2}{4\pi}a\wedge\mathrm{d}a
     - \frac{1}{4\pi} \sum_{\sigma\in \uparrow,\downarrow} \alpha_\sigma \wedge \mathrm{d}\alpha_\sigma +\frac{1}{2\pi}(a+A)\wedge\mathrm{d}(\alpha_\uparrow+\alpha_\downarrow)  + \frac{1}{2\pi}A_s\wedge\mathrm{d}(\alpha_\uparrow-\alpha_\downarrow)+\dots,\\
     &= \frac{2}{2\pi}a\wedge\mathrm{d}A + \frac{2}{4\pi}A\wedge\mathrm{d}A+\frac{2}{2\pi}A_s\wedge\mathrm{d}A_s+\dots,
    \label{eq:effective_action_superconductor}
\end{aligned}
\end{equation}
without residual topological order. The field theory in Eq.~\ref{eq:effective_action_superconductor} allows us to infer that the superconductor is topologically equivalent to a d+id superconductor in the BCS regime~\cite{SenthilMarstonFisher,Song2021}, which has a chiral central charge of $c_-=2$ and retains the spin quantum Hall response of the parent state.
  
\subsection{Gauge theory description of the doped IQH state}
This parallels anyon-superconductivity arguments, for the `confined' integer quantum Hall regime. Before starting, we should remark that the simplest picture of the IQH phase is obtained by doping Cooper pairs directly, which are the only cheap local excitation in the phase which circumvent an energy penalty due to the open spin gap. This naturally inherits the spin response and chiral central charge of $c_- = 2$ from the IQH state, matching the topology of the SC formed by doping the CSL. Close to the critical point, this may also be described using the redundant gauge theory description. 

To describe the IQH, note that the chargon field is in the Higgs phase $\langle \varphi\rangle  \neq 0$. Thus, charge may be seen as entering the system as vortices of $\varphi$, which by virtue of the CS term carry electric charge $2e$. This is naturally described in the dual vortex formulation~\cite{Dasgupta1981,Fisher1989}, whose extension to include the Chern-Simons term was used to analyze anyon superconductivity~\cite{LeeAnyonSuperconductivity1991}.
The boson current is expressed as $j^\mu_b(x) = 1/2\pi\epsilon^{\mu\nu\sigma} \partial_\nu \beta_\sigma(x)$:
\begin{equation}
\begin{aligned}
    \tilde{\mathcal{L}}_{\mathrm{QCP}}[A,A_s] &=\frac{2}{4\pi}  (a+A) \wedge \mathrm{d}(a+A) + \frac{2}{4\pi}   A_s \wedge \mathrm{d}A_s +\frac{1}{2\pi} a\wedge\mathrm{d}\beta \\
    +\;&|(\partial_\mu -i\beta ) \Phi |^2 
    + \tilde{r} |\Phi|^2 - \tilde{\lambda} |\Phi|^4 + \dots,
    \label{eq:dual_effective_action_CSL}
\end{aligned}
\end{equation}
where $\Phi$ is the vortex field and we have already integrated out the spinon fields.
Upon doping charges, we naturally find that there are $n_\Phi=\frac{\delta}{2}$ vortices per unit cell, which are at filling $\nu_\Phi=\frac{1}{2}$. Hence the vortices may naturally form a $k=2$ bosonic Laughlin state which can be expressed as
\begin{equation}
    \mathcal{L}_\Phi [\beta] = \frac{1}{2\pi}\beta\wedge \mathrm{d}\gamma - \frac{2}{4\pi}\gamma\wedge\mathrm{d}\gamma.
\end{equation}
Combined with the IQH parent state and upon inegrating out $\beta$, we again recover a chiral superconductor without residual topological order
\begin{equation}
    \mathcal{L}^\text{IQH}_\mathrm{doped}= \frac{2}{4\pi} A\wedge\mathrm{d}A+ \frac{2}{4\pi}   A_s \wedge \mathrm{d}A_s +\frac{2}{2\pi}a\wedge \mathrm{d}A
\end{equation}
As expected, the superconductor obtained this way retains both the spin-response and chiral central charge from the $\nu_{\text{tot.}}=2$ IQH state. Thus the superconductor reached from both sides of the transition have identical topological response.
\end{document}